\documentclass[]{aastex631}

\usepackage{graphicx}
\usepackage{hyperref}
\usepackage{comment}

\newcommand{\ang}{\text{\AA}}
\newcommand{\ts}{\textsuperscript}

\begin{document}

\title{HST-COS Transit Spectroscopy of KELT-20b: First Detection of Excess Far-ultraviolet Absorption From an Ultra-hot Jupiter}

\author[0000-0002-0786-7307]{Patrick R. Behr}
\affiliation{Department of Astrophysical and Planetary Sciences, University of Colorado Boulder\\ Boulder, CO 80309, USA}
\affiliation{Laboratory for Atmospheric and Space Physics,
Boulder, CO 80303, USA}

\author[0000-0002-1002-3674]{Kevin France}
\affiliation{Department of Astrophysical and Planetary Sciences, University of Colorado Boulder\\ Boulder, CO 80309, USA}
\affiliation{Laboratory for Atmospheric and Space Physics,
Boulder, CO 80303, USA}

\author[0000-0003-4426-9530]{Luca Fossati}
\affiliation{Space Research Institute,
Austrian Academy of Sciences, Austria}

\author{Tommi Koskinen}
\affiliation{University of Arizona,
Tuscon, AZ 85721, USA}

\author[0000-0002-1347-2600]{Patricio E. Cubillos}
\affiliation{Space Research Institute,
Austrian Academy of Sciences, Austria}
\affiliation{Osservatorio Astrofisico di Torino, 
Via Osservatorio 20, 
10025 Pino Torinese, Italy}

\author[0000-0002-4701-8916]{Arika Egan}
\affiliation{Applied Physics Laboratory, Johns Hopkins University,
Laurel, MD, 20723, USA}
\author[0000-0001-9207-0564]{P. Wilson Cauley}
\affiliation{Laboratory for Atmospheric and Space Physics
Boulder, CO 80303, USA}

\begin{abstract}

KELT-20 b is an ultra-hot Jupiter with an equilibrium temperature of $2260$ K orbiting a bright (V =7.6), fast-rotating ($v\sin{i}$=117 km s$^{-1}$) A2 V star. The atmosphere of KELT-20 b has been studied extensively via transmission spectroscopy at optical wavelengths, showing strong hydrogen absorption as well as metals including \ion{Na}{1}, \ion{Ca}{2}, \ion{Fe}{1}, \ion{Fe}{2}, \ion{Mg}{1}, \ion{Si}{1} and \ion{Cr}{2}. The atmospheric and ionization conditions of this planet may differ from Jupiter-mass exoplanets due to the relatively weak extreme-ultraviolet radiation from its host star, as the stellar dynamo that generates chromospheric and coronal activity is thought to shut down at spectral types earlier than A4. We present the first spectroscopic observations of KELT-20 b in the far-ultraviolet using the Hubble Space Telescope Cosmic Origins Spectrograph, searching for previously undetected low-ionization and neutral atoms in the upper atmosphere. We find that the FUV transit depth increases with decreasing wavelengths, from $1.88\pm0.04$\% at 1600--1760 \ang\ to $2.28\pm0.04$\% at 1410--1570 \ang, yielding planetary radii of $0.1139\pm0.06$ $R_*$ and $0.1222\pm0.07$ $R_*$, respectively. We report tentative detections of \ion{Fe}{2} and \ion{N}{1} at $2.4\sigma$ each, and non-detections of \ion{C}{1}, \ion{S}{1}, \ion{Al}{2}, and \ion{Si}{2}. We find no evidence for molecular absorption from CO or H$_2$ and no sign of hydrodynamic escape.

\end{abstract}


\section{Introduction} \label{sec:intro}

Ultra-hot Jupiters (UHJs) are broadly defined as planets with dayside temperatures of $T_{\rm{eq}} > 2000$ K, radii of R$_{\text{p}} > \text{R}_{\text{Jupiter}}$, and periods of $P<5$ days \citep{parmentier_thermal_2018}. Many UHJs are found orbiting hot early-type stars with photospheres emitting into the far-ultraviolet (FUV; $912 < \lambda < 1700$ \ang). However, stellar chromospheric and coronal activity, which produce the higher energy X-ray and extreme-UV (in combination, XUV; $\lambda<912$ \ang) photons responsible for heating the upper atmosphere and powering escape on most short period planets, decreases with increasing effective temperature and eventually shuts off entirely for temperatures $T_{\text{eff}}\gtrsim8300$ K \citep{simon_limits_2002,neff_o_2008}. Planets orbiting these early-type stars are therefore some of the most FUV-irradiated planets known but may experience relatively little XUV flux \citep{fossati_extreme-ultraviolet_2018}, providing a unique opportunity to characterize planetary atmospheres in FUV-dominated environments.

UV photons are absorbed high in the planetary atmosphere and many atmospheric constituents have large UV absorption cross-sections. This makes UV transit spectroscopy particularly effective for studying the atmospheric composition, energy balance, and mass loss rates from the upper atmosphere of giant planets \citep{vidal-madjar_detection_2004,koskinen_2013_hd209458b,cubillos_2020_hd209458b,huang_2023_hydrodynamic,dos-santos_2023_escape}. However, FUV observations of planets around late-type stars can be difficult due to low levels of FUV flux and chromospheric activity driving UV flares and short-timescale variability \citep{feinstein_hst_2024}. In the case of hot, early-type host stars, these complications are minimized or even eliminated due to the strong photospheric emission and lack of chromospheric activity; these stars provide a very bright and temporally stable FUV background against which the many atoms, ions, and molecules with resonant transitions in the FUV can be observed. In this paper we present the first FUV transit spectroscopy of an UHJ.

KELT-20 b (MASCARA-2 b) has a temperature of $T_{\rm{eq}}=2260$ K, a period of $P=3.47$ days, and orbits a magnitude $\rm{V}=7.6$, fast-rotating ($v\sin{i}$=117 km s$^{-1}$) A2 star with $T_{\rm{eff}}$ between 8700--9000 K \citep{lund_kelt-20b_2017,talens_mascara-2_2018}. Previous transit observations of KELT-20 b using optical spectroscopy have shown deep H Balmer lines as well as evidence of atmospheric metals including \ion{Na}{1}, \ion{Ca}{2}, \ion{Fe}{1}, \ion{Fe}{2}, \ion{Mg}{1}, \ion{Si}{1} and \ion{Cr}{2} \citep{casasayas-barris_atmospheric_2019,nugroho_searching_2020,hoeijmakers_high-resolution_2020,cont_silicon_2022,fossati_gaps_2023,stangret_gaps_2024}. Infrared observations during secondary eclipse have shown evidence of H$_2$O and CO in emission from the dayside \citep{fu_strong_2022}. We have obtained FUV transit spectroscopy of KELT-20 b using the Hubble Space Telescope Cosmic Origins Spectrograph (HST-COS). FUV spectroscopic observations probe the many neutral and low-ionization metals and molecules (e.g., \ion{S}{1}, \ion{C}{1}, \ion{N}{1}, \ion{Si}{2}, \ion{Al}{2}, \ion{Fe}{2}, and CO) which may be present in the upper atmosphere of KELT-20 b.

We describe the HST-COS observations and data reduction in Section \ref{sec:observations}, the transit analysis and modeling in Section \ref{sec:transit-analysis}, discuss results in Section \ref{sec:results}, and provide a summary of our results in Section \ref{sec:summary}.

\section{Observations}
\label{sec:observations}

We observed one transit of KELT-20 b over 5 orbits of HST with the COS G160M grating centered at 1589 \ang, totaling 11,952 seconds of exposure (HST cycle 30, program ID 17156; PI Cauley). The G160M grating provides moderate spectral resolving power of R=13,000--20,000 and spans wavelengths from 1403--1775 \ang\ in this configuration. Information for each orbit can be found in Table \ref{tab:obslist}. 

\begin{deluxetable}{ccc}[!ht]
\label{tab:obslist}
\tablecaption{HST-COS observations of KELT-20 b transit}
\tablehead{\colhead{Dataset} & \colhead{Observation Start} & \colhead{Exposure Time} \\ 
\colhead{} & \colhead{(UTC)} & \colhead{(s)} } 
\startdata
    LEXB02CZQ & 2023-09-11 09:27:25 & 2052 \\
    LEXB02D5Q & 2023-09-11 10:54:36 & 2475 \\
    LEXB02DBQ & 2023-09-11 12:29:40 & 2475 \\
    LEXB02DJQ & 2023-09-11 14:04:44 & 2475 \\
    LEXB02DQQ & 2023-09-11 15:39:48 & 2475 \\
\enddata
\end{deluxetable}

We also observed four transits of KELT-20 b with the Colorado Ultraviolet Transit Experiment CubeSat (CUTE; \citet{france_colorado_2023}) between July 12\ts{th} and October 21\ts{st}, 2023. The CUTE CubeSat conducts spectroscopic transit observations of short-period giant planets in the near-UV (NUV) spectral band between 2479--3306 \ang \citep{cute_wasp-189_2023,cute_kelt-9_2024}. All CUTE exposures are 5 minutes in duration, totaling approximately 31 hours per visit, and cover roughly -0.2--0.2 of the planet's orbital phase.

\subsection{HST-COS Data reduction}
\label{sec:data-reduction}
\begin{figure}[!ht]
    \centering
    \includegraphics[width=\linewidth]{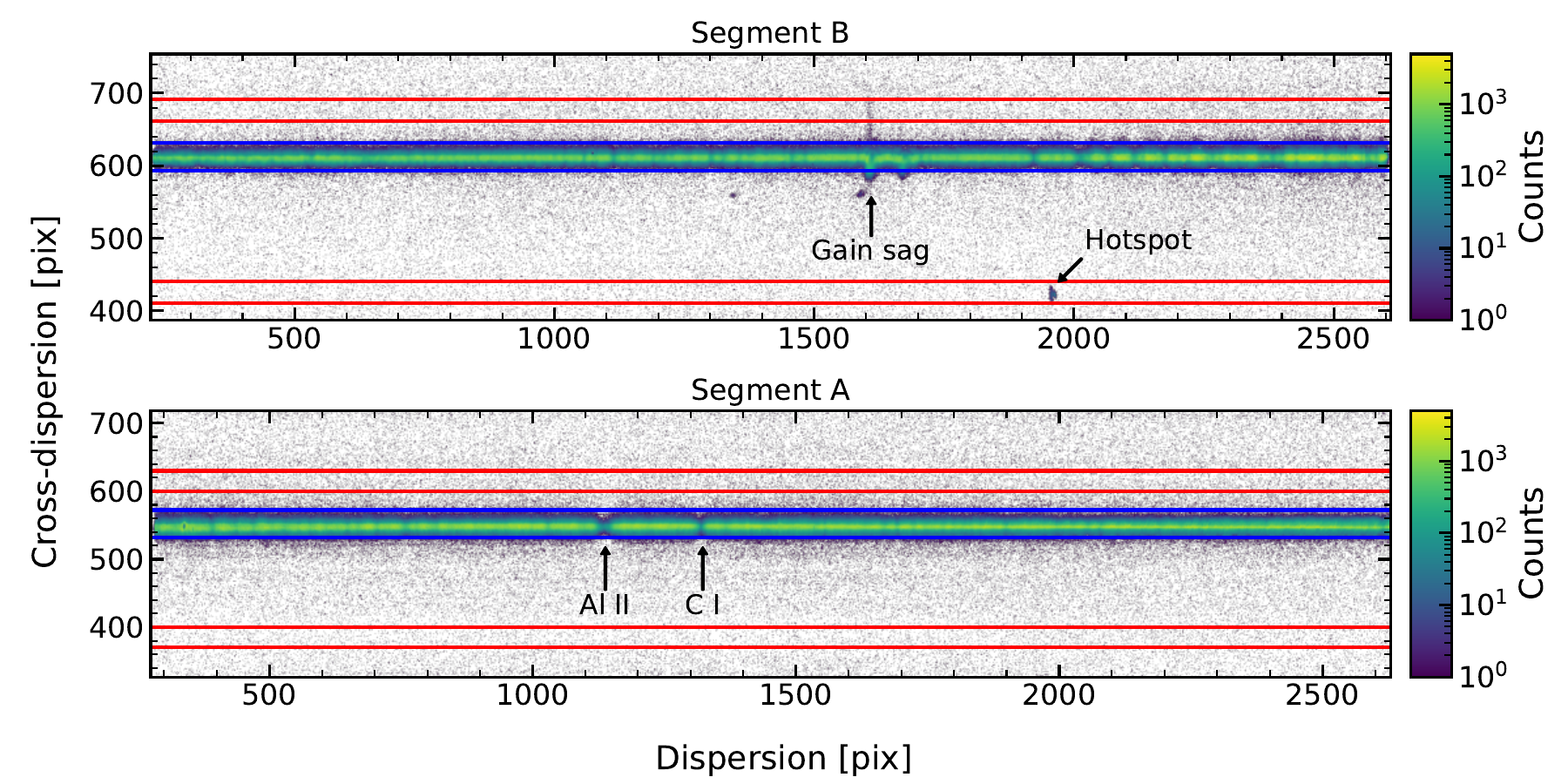}
    \caption{Out-of-transit 2D spectrum of KELT-20 before data reduction. Both axes are in units of detector pixels. Data are binned to 6 pixels in the dispersion direction and unbinned in the cross-dispersion direction. Denoted in the image are the strong \ion{Al}{2} ($\lambda1671$ \ang) and \ion{C}{1} ($\lambda1657$ \ang) absorption features as well as detector locations which suffer from gain-sag and background hotspots (see \S \ref{sec:data-reduction}). The two background extraction regions are shown with horizontal red lines. The spectrum extraction region is shown with horizontal blue lines.}
    \label{fig:spec2d}
\end{figure}

\begin{figure}
    \centering
    \includegraphics[width=\linewidth]{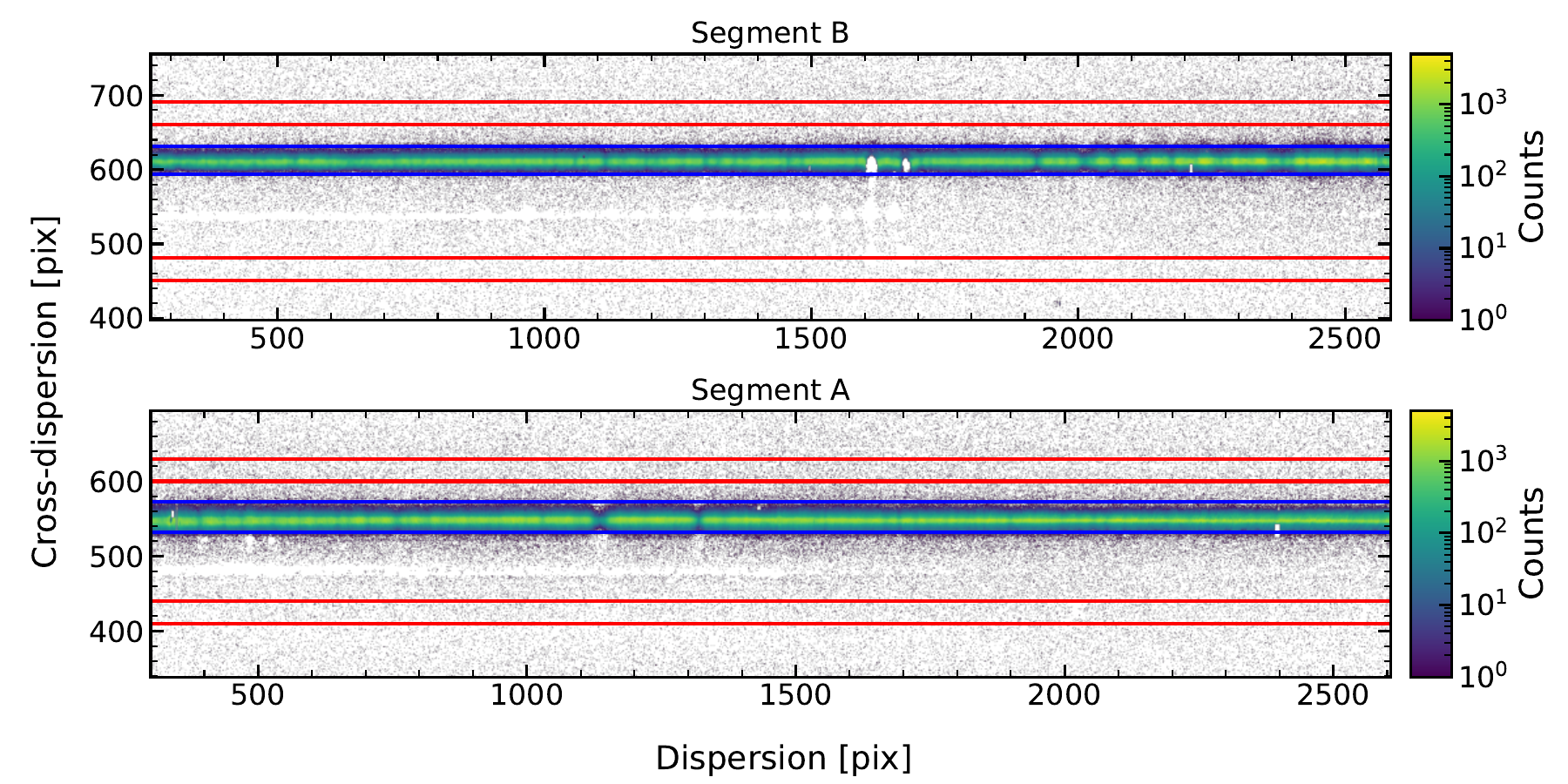}
    \caption{Same as Figure \ref{fig:spec2d} but after data reduction. Events with SDQ flags and low PHA have been removed. The hotspot in Segment B is not fully removed so we shifted the lower background region upwards 40 pixels in the cross-dispersion direction direction.}
    \label{fig:spec2d_clean}
\end{figure}

The COS FUV detector is a photon-counting microchannel plate (MCP) sensor split into two segments, A and B. Segment A spans wavelengths between 1595--1768 \ang\ and segment B wavelengths between 1403--1576 \ang. We perform our analyses using the calibrated photon event list \texttt{corrtag} files. The \texttt{corrtag} file has been processed with the \texttt{calcos} version 3.4.7 pipeline, which generates time-tagged photon events with X and Y detector locations corrected for thermal and geometric distortions as well as time of arrival, wavelength, pulse height amplitude (PHA), and data quality (DQ) flags. During data reduction, we first identify and remove any data that have a serious data quality (SDQ) flag. For the COS FUV detector, these are flags 2 (hot spot), 8 (poorly calibrated region), 16 (low response region), 128 (pixel out-of-bounds), and 8192 (gain-sag hole). We also remove any events with data quality flags of 512, representing a PHA less than 2 or greater than 23 (default values from \texttt{calcos}). In our observations, events with a PHA less than 2 primarily result from a ``gain sag halo''---an area of low gain surrounding locations flagged for gain sag but not flagged as SDQ events. In total, the data cleaning process removed $\sim3\%$ of the data from segment A in each orbit and $\sim5\%$ of the data from segment B in each orbit. The raw 2D spectrum from the \texttt{corrtag} events list is shown in Figure \ref{fig:spec2d} and the post-processing 2D spectrum is shown in Figure \ref{fig:spec2d_clean}.

We performed background subtraction at the \texttt{corrtag} level following the procedure outlined in the two-zone extraction method described in the \texttt{BACKCORR} process in the COS Data Handbook \S 3.4.20. The background region is composed of two rectangular sections (one above and below the stellar spectral trace) with heights of 41 pixels and width spanning the entire detector. Inspection of the 2D spectra show a localized hotspot in the lower background region of the detector B segment (annotated in Figure \ref{fig:spec2d}) in every exposure. The hotspot is not fully removed during the data quality processing; in order to avoid over-subtraction, we moved the lower background region of detector segment B up 40 pixels in the cross-dispersion direction. The background and source regions are then collapsed in the cross-dispersion direction to obtain 1D spectra in units of counts \ang$^{-1}$. The two background regions are summed and scaled by the ratio of $H_{\rm{src}}/(2H_{\rm{bkg}})$, where $H_{\rm{src}}$ is the height in pixels of the spectrum extraction region and $H_{\rm{bkg}}$ the height of the background regions. The scaled 1D background spectrum is then subtracted from the 1D source spectrum. Finally, the background-subtracted spectrum is weighted by the exposure time to create a spectrum in counts s$^{-1}$ \ang$^{-1}$ which is used for analysis.

\subsubsection{Instrument systematics}
\label{sec:instrument-effects}
\begin{figure}[!ht]
    \centering
    \includegraphics[width=0.9\linewidth]{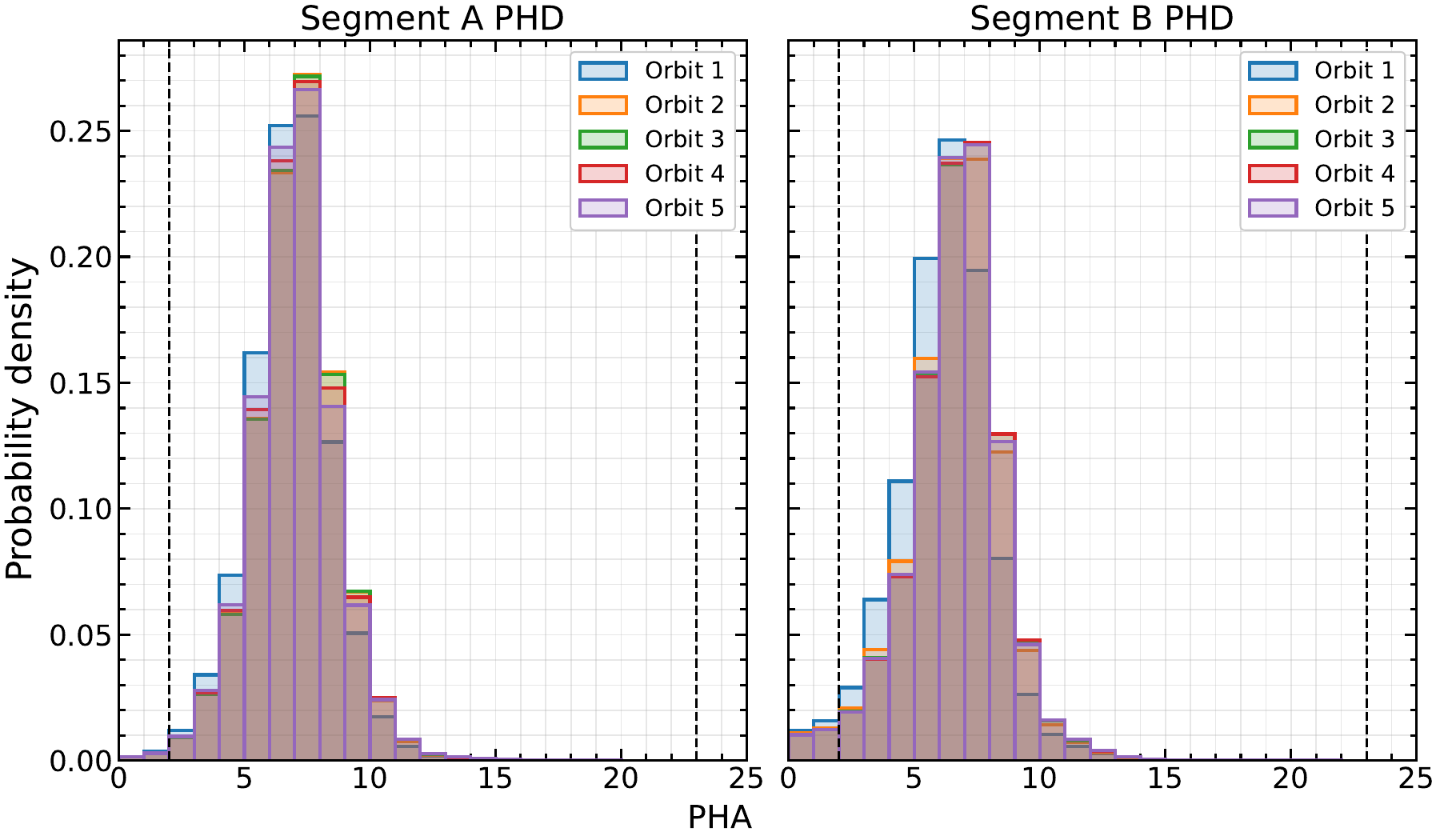}
    \caption{Pulse height distribution (PHD) for each COS detector segment. Distributions are shown as density plots to account for different exposure times. Dashed vertical line show the upper and lower PHA limit set in \texttt{calcos}.}
    \label{fig:detector-PHD}
\end{figure}

\begin{figure}[!ht]
    \centering
    \includegraphics[width=0.9\linewidth]{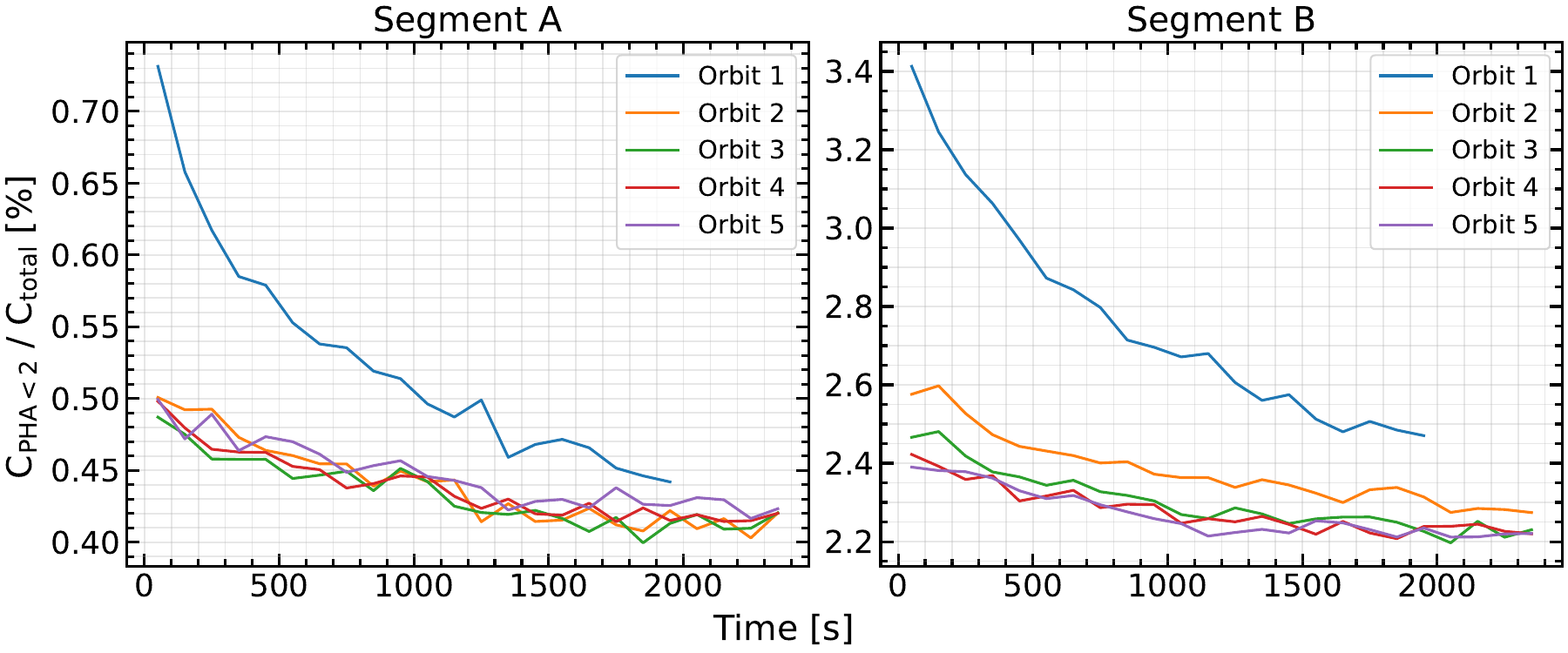}
    \caption{Time series of low PHA events in each detector segment for each orbit. Data are shown as the ratio of counts with $\rm{PHA}<2$ to the total counts in each bin. Orbit 1 shows excessive low PHA counts in both detector segments.}
    \label{fig:pha-light-curve}
\end{figure}

Transit observations with the HST Space Telescope Imaging Spectrograph (HST-STIS) are known to exhibit large and periodic systematic effects due to the telescope's thermal cycle and pointing stability which can drive targets out of the slit and change the point-spread-function (PSF) \citep{brown_hubble_2001,sing_hubble_2019}. Thanks to the large open aperture of HST-COS---2.5'' compared to the commonly used 0.2'' slit of STIS---our observations are not subject to loss of light from small pointing discrepancies. However, we may still be sensitive to PSF changes and detector gain sag. Both detector segments suffer from excessive amounts of low PHA events during orbit 1 but segment B is more impacted because of the two highly gain-sagged regions (pointed out in Figure \ref{fig:spec2d}) due to repeated exposure to bright features (notably \ion{H}{1} Ly$\alpha$) which are not present in segment A. The gain-sagged regions also suffer from un-corrected y-walk which causes them to be shifted downward in the cross-dispersion direction. Figure \ref{fig:detector-PHD} shows the pulse height distribution (PHD) for both detector segments and Figure \ref{fig:pha-light-curve} shows the time series of events with $\rm{PHA}<2$.

\begin{figure}[!ht]
    \centering
    \includegraphics[width=\linewidth]{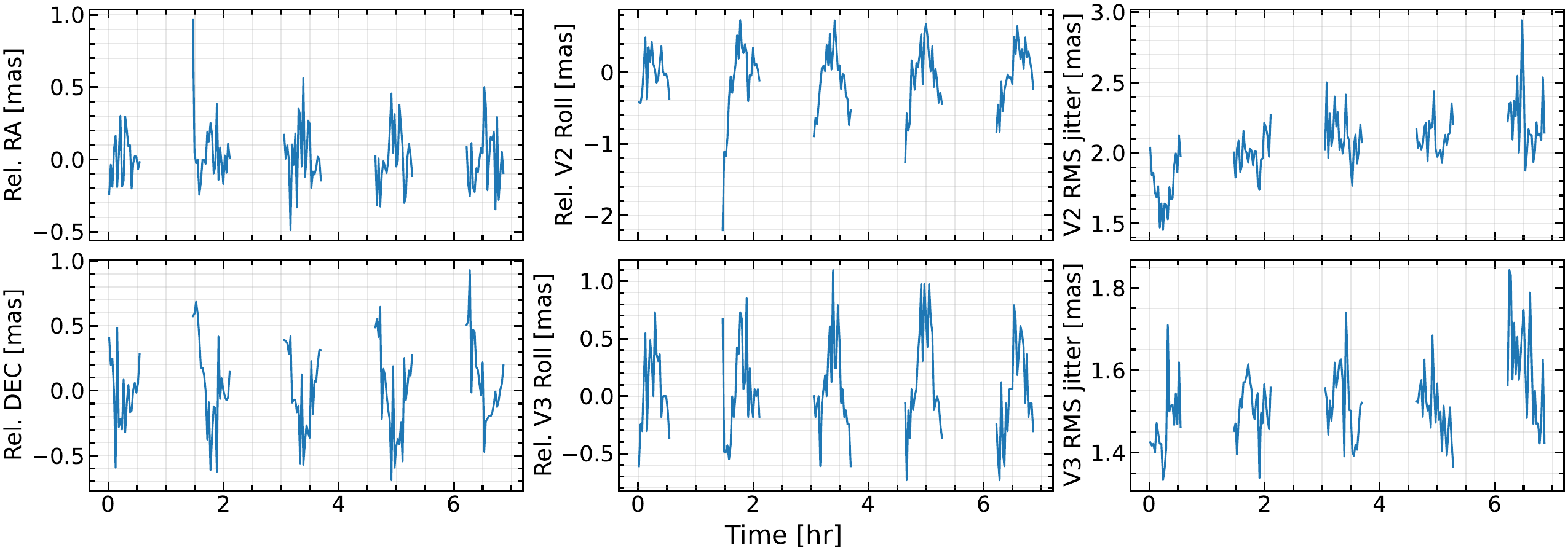}
    \caption{RA, DEC, roll, and RMS jitter during each orbit. RA, DEC, and roll are shown relative to the median value across all orbits.}
    \label{fig:pointing}
\end{figure}

Next, we examine the pointing stability during each orbit. Engineering data were extracted from the \texttt{jit} files and processed similarly to \citet{sing_hubble_2019}. Figure \ref{fig:pointing} shows the R.A., Dec, and roll relative to their median values across all orbits as well as the RMS jitter for each orbit. Pointing was stable with a peak-to-trough amplitude of only 1 mas so we do not expect any significant systematic effects due to pointing instability.

\begin{figure}[!ht]
    \centering
    \includegraphics[width=0.9\linewidth]{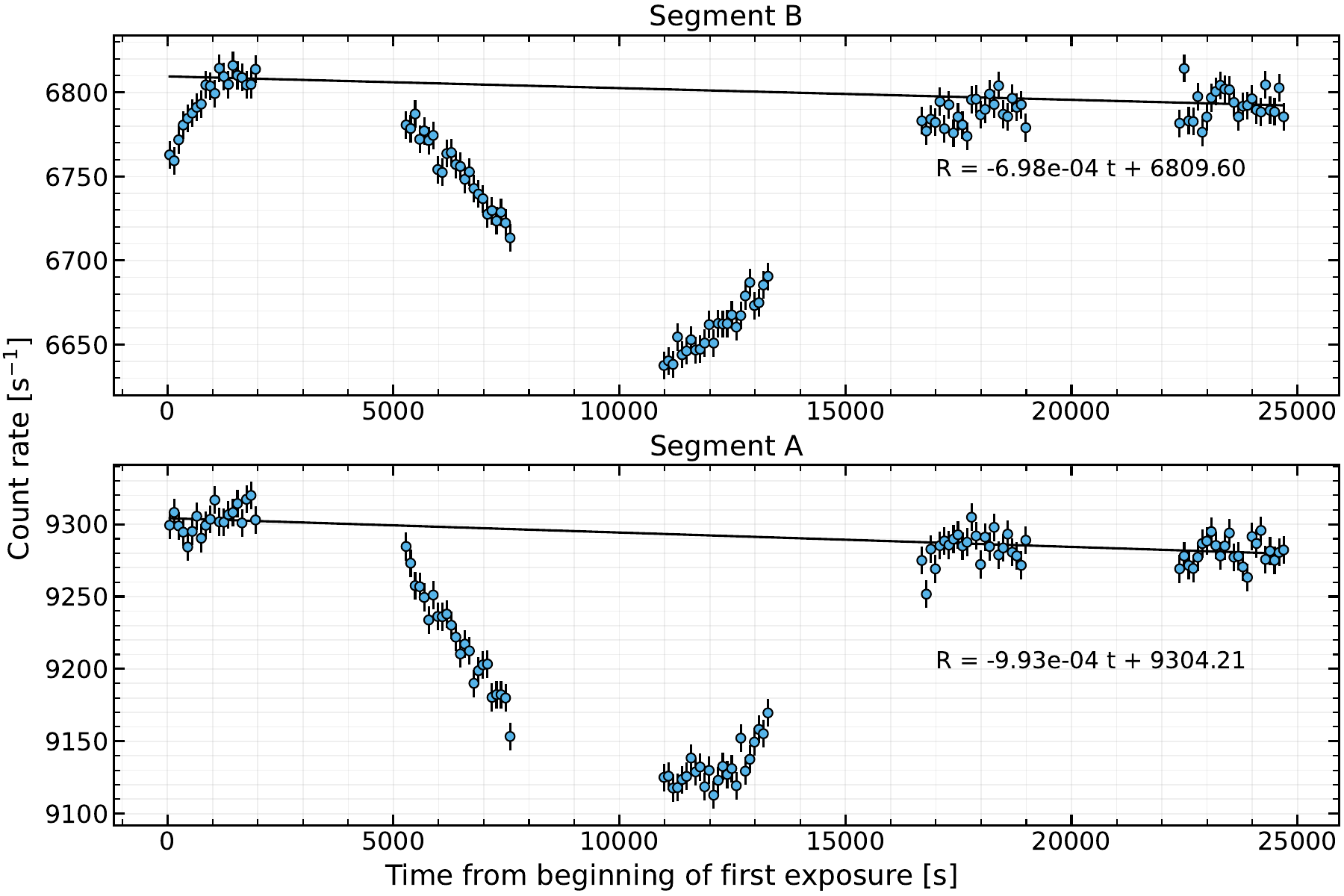}
    \caption{Light curve of the KELT-20 b transit after removing all SDQ and low-PHA events. The light curves show an transit-level decrease in count rate over the five orbits. The solid black lines are linear fits to the count rates in orbit 1 and orbit 5. Over the full duration of observation, the count rate decreases by 7.3 s$^{-1}$ in segment B and 28.1 s$^{-1}$ in segment A. These trends are assumed to be due to instrument systematics as seen in other HST transit observations. The data are divided by the linear fits to obtain the final, fully-corrected light curves.}
    \label{fig:linear-trend}
\end{figure}

\begin{figure}[!ht]
    \centering
    \includegraphics[width=0.9\textwidth]{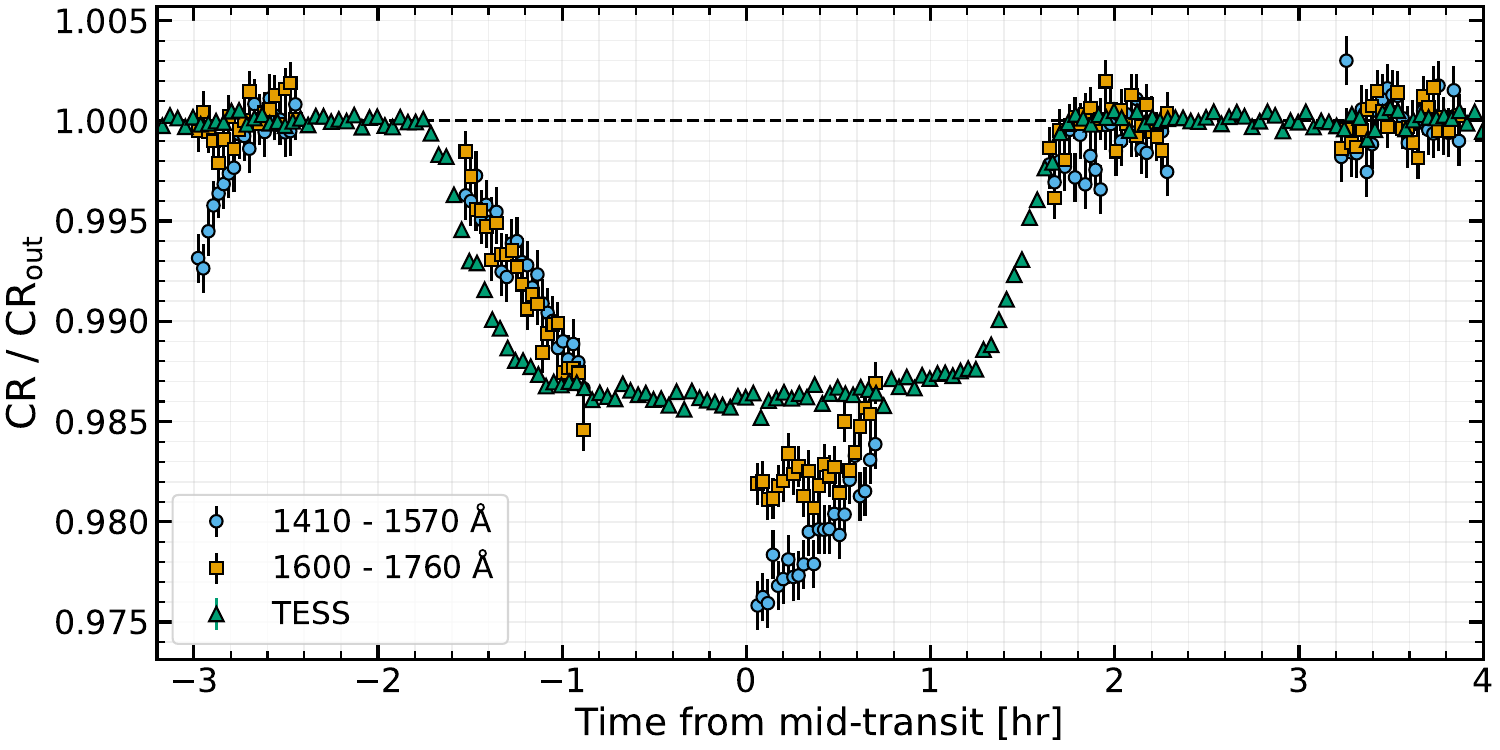}
    \caption{KELT-20 b transit light curve after removing events with SDQ flags or out-of-bounds PHA. The TESS optical light curve is plotted with green triangles for comparison.}
    \label{fig:light-curve}
\end{figure}

After removing SDQs, low PHA events, and performing background subtraction, we still see a linear trend of decreasing count rate across the five orbits with 0.2\% difference between orbit 1 and orbit 5. The trend is assumed to be due to orbit-related instrument systematics as seen in other HST instruments \citep[e.g,][]{brown_hubble_2001,sing_hubble_2019}. We correct this by fitting a line between orbit 1 (excluding the first 1 ks due to the low PHA events) and all of orbit 5 then dividing all data by the resulting trend line. Figure \ref{fig:linear-trend} shows the light curves for both detector segments before removing the linear trend. Figure \ref{fig:light-curve} shows the trend-corrected light curve compared with the TESS optical light curve\footnote{Phase folded light curve from obsid tess2021175071901-s0040-0000000069679391-0211-s retrieved from MAST}.

\subsection{CUTE data reduction}
\label{sec:cute-reduction}
CUTE data were processed using the CUTE autonomous data reduction pipeline (\texttt{CONTROL}; \citet{sreejith_autonomous_2022}), which carries out both background subtraction and spectral extraction to produce 1D spectra through the following process: each image is bias-subtracted using a single master bias frame. On each frame, the spectral extraction region is identified, and an equally sized region below the spectrum is used for background subtraction, leaving a background-subtracted 2D spectral trace. This is then summed along the cross-dispersion axis to create a 1D spectrum. The \texttt{CONTROL}-produced 1D spectra are then summed along the dispersion axis to produce a broadband NUV ``white light'' time series as in \citet{egan_colorado_2024}. The white light curve is the sum of stellar flux and a 12 parameter model to account for scattered light and instrument systematics. To improve the signal-to-noise of the final white light curve, we interpolated each transit observation onto a common time grid and coadded the three visits, weighted by the uncertainty of each observation. The coadded light curve was then binned to a temporal cadence of 90 minutes.

We fit the coadded light curve using the \texttt{PyTransit} RoadRunner model. During the fitting process, the planet-to-star radius ratio was allowed to vary while the remainder of the orbital parameters were held fixed to literature values. Limb-darkening was implemented using the non-linear limb-darkening law with coefficients fixed to values retrieved from the web browser \texttt{ExoCTK} limb-darkening calculator version 1.2.6\footnote{\url{https://exoctk.stsci.edu/limb_darkening}} \citep{matthew_bourque_2021_4556063}.

\begin{figure}[!ht]
    \centering
    \includegraphics[width=0.9\linewidth]{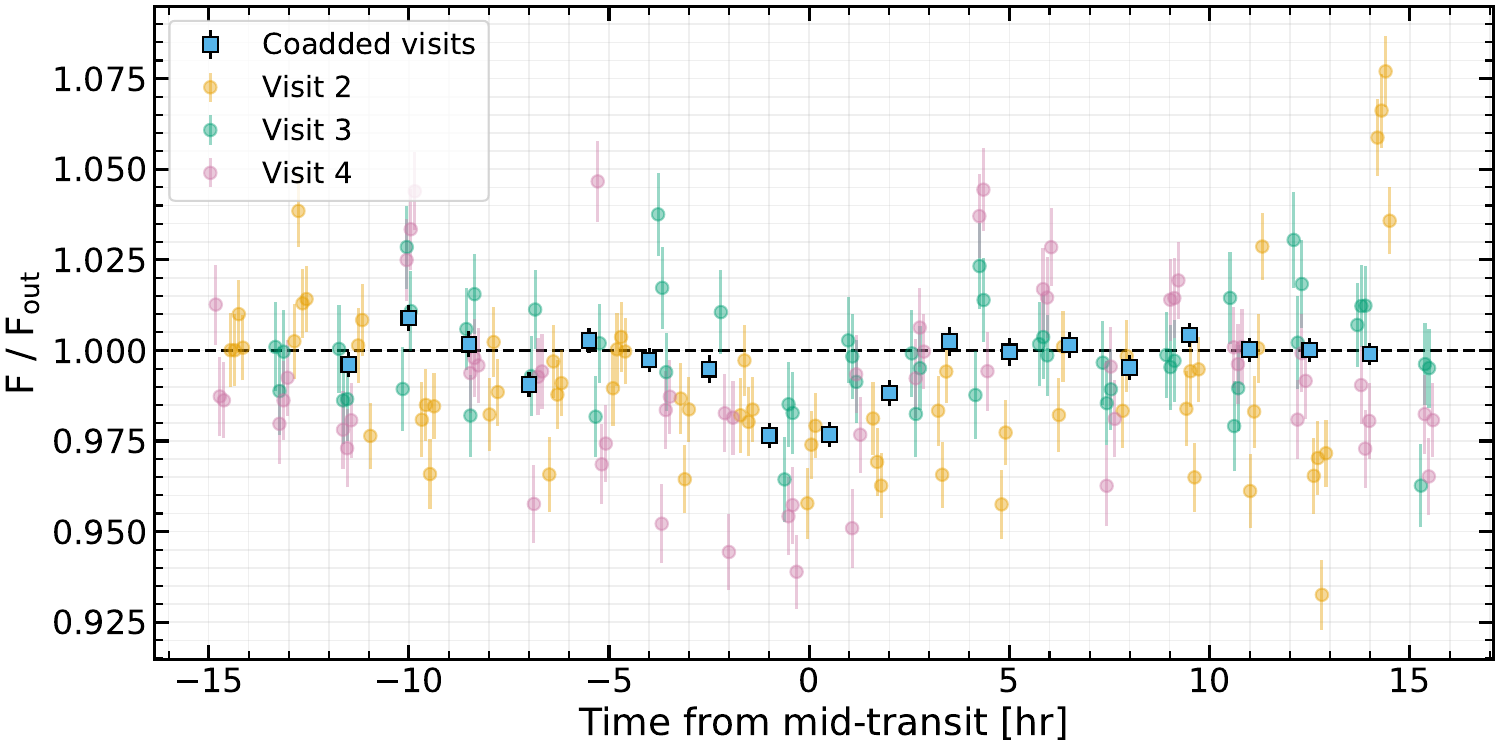}
    \caption{CUTE broadband NUV light curves for visits 2--4. All data are shown after being processed with the \texttt{CONTROL} pipeline. Individual visits are shown with semi-transparent circles at full temporal resolution. The final coadded light curve is shown in solid blue squares and is binned to a 90 minute cadence.}
    \label{fig:cute-light-curve}
\end{figure}

\section{Transit analysis}
\label{sec:transit-analysis}
\subsection{Transit model}
\label{sec:model}
We fit the HST-COS transit light curve using the \texttt{PyTransit} version 2.6.4 RoadRunner model\footnote{\url{https://pytransit.readthedocs.io/en/latest/}} \citep{parviainen_pytransit_2015} and performed a Markov chain Monte Carlo (MCMC) analysis using \texttt{emcee} \citep{foreman-mackey_emcee_2013} to retrieve the best-fit parameters. Our MCMC varies two parameters: planet-to-star radius and time of mid-transit. All other orbital parameters are fixed to previously published results, shown in Table \ref{tab:orbit-params}, and the limb darkening coefficients are fixed to values retrieved from the \citet{espinoza_limb_2015} \texttt{limb-darkening} code (see \S \ref{sec:limb-darkening}). The model was run with a number of walkers equal to eight times the number of free parameters for 20,000 steps with a burn-in of 2,000 steps.

\begin{deluxetable}{CcCc}[!h]
\label{tab:orbit-params}
\tablewidth{\textwidth}
\tablecaption{Adopted parameters for the KELT-20 system}
\tablehead{\colhead{Parameter} & \colhead{Description} & \colhead{Value}    & \colhead{Source}}
\startdata
    T_{\rm{eff}} &   Stellar effective temperature [K]   &   8800    &   This work\\
    M_* &   Stellar mass [M$_\odot$]    &   1.89    &   \citet{talens_mascara-2_2018}\\
    R_* &   Stellar radius [R$_\odot$]  &   1.60    &   \citet{talens_mascara-2_2018}\\
    \gamma& Systemic velocity [km s$^{-1}$] &   -21.07  &   \citet{talens_mascara-2_2018}\\
    \log{(g)}   &   Surface gravity [cm s$^{-2}$]   &   4.31    &   \citet{talens_mascara-2_2018}\\
    \hline
    M_p &   Planet mass [M$_{Jup}$]   &   < 3.382   &   \citet{lund_kelt-20b_2017}\\
    R_p &   Planet radius [R$_{Jup}$]   &   1.741   &   \citet{lund_kelt-20b_2017}\\
    a  &   Semi-major axis [R$_*$]    &    7.42 &   \citet{lund_kelt-20b_2017}\\
    e  &   Eccentricity    &   0.00 &   \citet{lund_kelt-20b_2017}\\
    i  &   Inclination [deg] &   86.12  &   \citet{lund_kelt-20b_2017}\\
    p   &   Period [days]   &   3.47410024  &   \citet{ivshina_tess_2022}\\
    t_c &   Time of conjunction [BJD] &   2458312.58564   &   \citet{ivshina_tess_2022}\\
\enddata
\tablecomments{Stellar effective temperature is based on the PHOENIX model discussed in \S \ref{sec:limb-darkening}}
\end{deluxetable}

\subsubsection{Stellar limb-darkening}
\label{sec:limb-darkening}
Stellar limb darkening, the decrease in observed intensity at the stellar edge as compared to the center, plays a role in determining the shape of an exoplanet transit. It is important to include limb darkening in a transit model so as not to misinterpret changes in stellar intensity as differences in planetary parameters---particularly $R_p/R_*$ which can be significantly overestimated if limb-darkening is not included. It is typical to express the shape of the stellar intensity curve as a polynomial in terms of $\mu=\cos{\theta}$, where $\theta$ is the angle between the line of sight and the normal to the stellar surface. Most commonly used are the quadratic (Eq. \ref{eq:quad}) and non-linear (Eq. \ref{eq:non-lin}) laws \citep[and references therein]{claret_new_2000}:

\begin{equation}
\label{eq:quad}
    \frac{I(\mu)}{I(1)} = 1-u_1(1-\mu)-u_2(1-\mu)^2
\end{equation}

\begin{equation}
\label{eq:non-lin}
    \frac{I(\mu)}{I(1)} = 1-\sum_{n=1}^4c_n(1-\mu^{n/2})
\end{equation}

where $u_n$ and $c_n$ are referred to as the limb darkening coefficients (LDCs). 

\begin{figure}[!h]
    \centering
    \includegraphics[width=\linewidth]{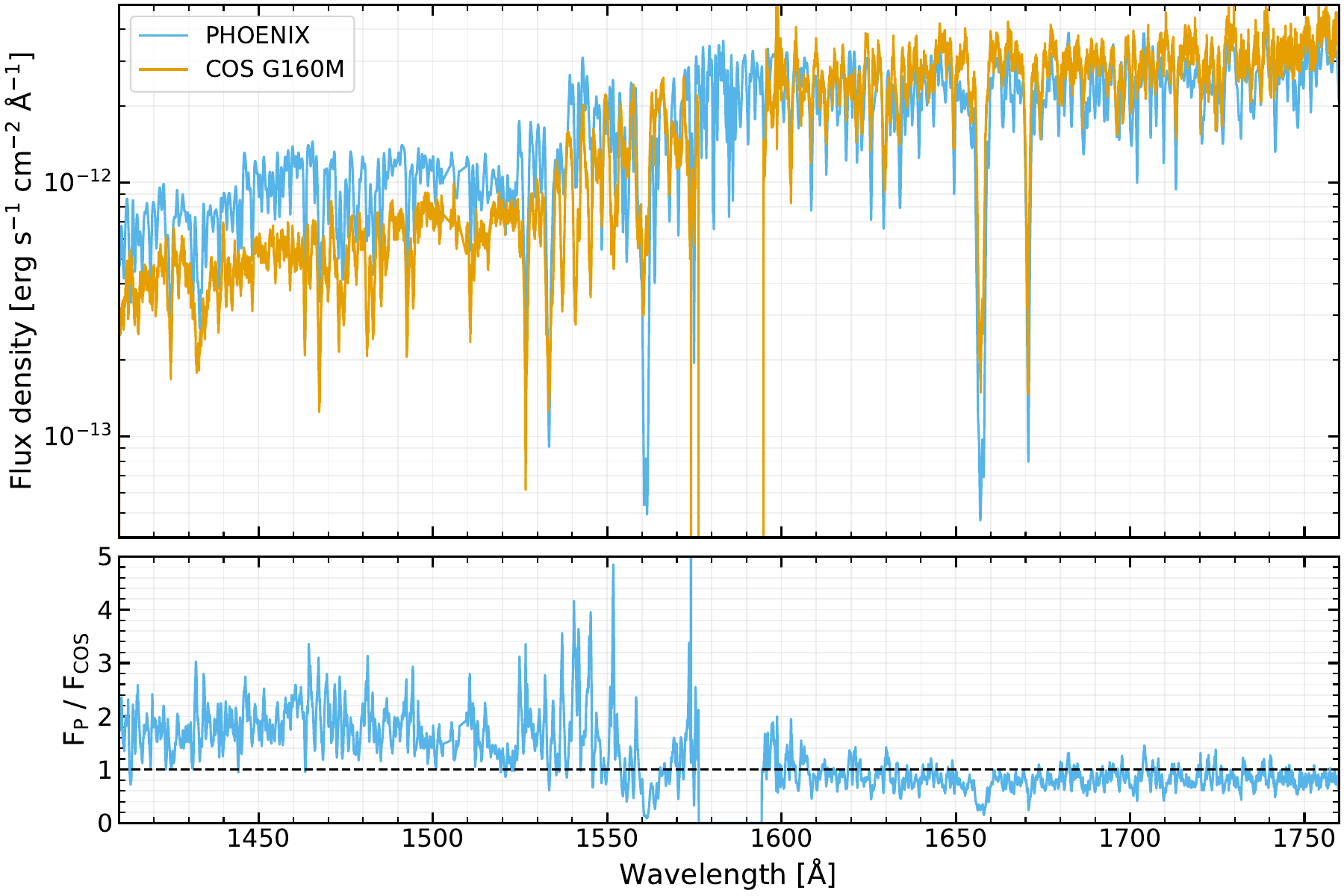}
    \caption{\textbf{Top:} Overplot of the averaged PHOENIX model, convolved to the spectral resolution of COS G160M, with the coadded out-of-transit COS spectra. We note that because the PHOENIX model is in units of flux density, we have plotted the COS data using the flux-calibrated \texttt{x1d} files created by the \texttt{calcos} pipeline. \textbf{Bottom:} Ratio of flux in the PHOENIX model to the out-of-transit COS spectrum.}
    \label{fig:phoenix}
\end{figure}

\begin{figure}[!ht]
    \centering
    \includegraphics[width=0.8\linewidth]{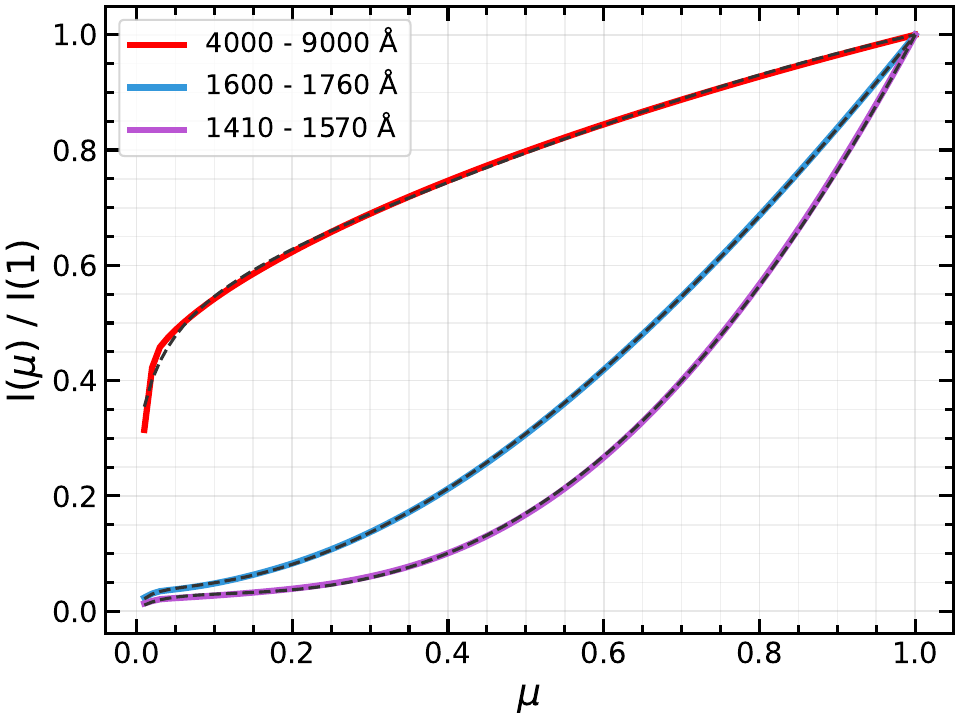}
    \caption{Stellar intensity profile obtained from the averaged PHOENIX model with $T_{\rm{eff}}\sim8800$ K and $\log(g)\sim4.25$ cm s$^{-2}$. The three intensity profiles shown are: visible/near-IR as observed with Kepler (red), COS G160M segment A (blue), and COS G160M segment B (purple). Dashed gray lines over each intensity profile represent the best fit model using the non-linear limb darkening law with coefficients retrieved from \texttt{limb-darkening}.}
    \label{fig:stellar-intensity}
\end{figure}

These laws have been well studied, used extensively, and shown to produce acceptable results for many different stellar types and transit parameters. There have been several studies investigating the various biases introduced by different methods of handling limb darkening. Of particular importance are the effects of leaving LDCs as free parameters versus fixing them to model or tabulated values \citep{espinoza_limb_2016,patel_empirical_2022}, and allowing non-physical solutions to fully sample a parameter space versus allowing only physical solutions \citep{kipping_efficient_2013,coulombe_biases_2024}. However, we note that these studies are typically focused on stars cooler than KELT-20 and at visible and IR wavelengths (i.e., the bandpasses of Kepler, CoRoT, TESS, and JWST) and have not been verified at FUV wavelengths. In order to verify that standard limb darkening laws are acceptable for our analysis, we obtained two PHOENIX models \citep{husser_new_2013} with temperatures of $T_{\rm{eff}}=8800$ K and $\log(g)$ of 4.00 and 4.50 cm s$^{-2}$. We average the models to obtain a single model with approximate values of $T_{\rm{eff}}=8800$ K and $\log(g)\sim4.25$ cm s$^{-2}$, which is consistent with previously published values \citep{lund_kelt-20b_2017,talens_mascara-2_2018}. We find good agreement between the averaged PHOENIX model and our out-of-transit HST-COS spectrum, as shown in Figure \ref{fig:phoenix}. We note, however, that the PHOENIX model over-predicts the absorption strength in the \ion{C}{1} lines at 1560 \ang\ and 1657 \ang, and slightly over-predicts the continuum in segment B. We then used the averaged PHOENIX spectrum to obtain the stellar intensity profile, $I(\mu)$, using the \texttt{limb-darkening} code of \citet{espinoza_limb_2015}. Figure \ref{fig:stellar-intensity} shows the stellar intensity profile as a function of viewing angle retrieved from our averaged PHOENIX models. The shape of the intensity profile in the FUV differs greatly from that in the visible region; however, the profile is still monotonically decreasing, as expected because the temperature of KELT-20 is high enough that FUV wavelengths are still probing the photosphere.

We find that the non-linear law provides an excellent fit to the PHOENIX intensity profile over the entire G160M bandpass. The quadratic law provides an equally good fit for segment A (1600--1800 \ang) but results in an increase in intensity towards the stellar limb for segment B (1400--1600 \ang), which we assume to be non-physical. Therefore, we opt to use the non-linear limb darkening law (Eq. \ref{eq:non-lin}) in the \texttt{PyTransit} model. 

Finally, we address the decision on whether to leave the LDCs as free parameters or to keep them fixed to the best-fit values from the PHOENIX model. We performed three light curve fits, allowing the LDCs to vary with uniform priors, vary with Gaussian priors centered on the values retrieved from \texttt{limb-darkening}, or remain fixed to the values from \texttt{limb-darkening}. Using the non-linear law with uniform priors we are unable to adequately constrain the LDCs, while using Gaussian priors provides well constrained LDCs but results in non-physical intensity profiles. We thus choose to keep the LDCs fixed to the values retrieved from \texttt{limb-darkening}.

\subsubsection{Rossiter-McLaughlin and CLV effects}
\label{sec:RM-CLV}
\begin{figure}[!ht]
    \centering
    \includegraphics[width=0.8\linewidth]{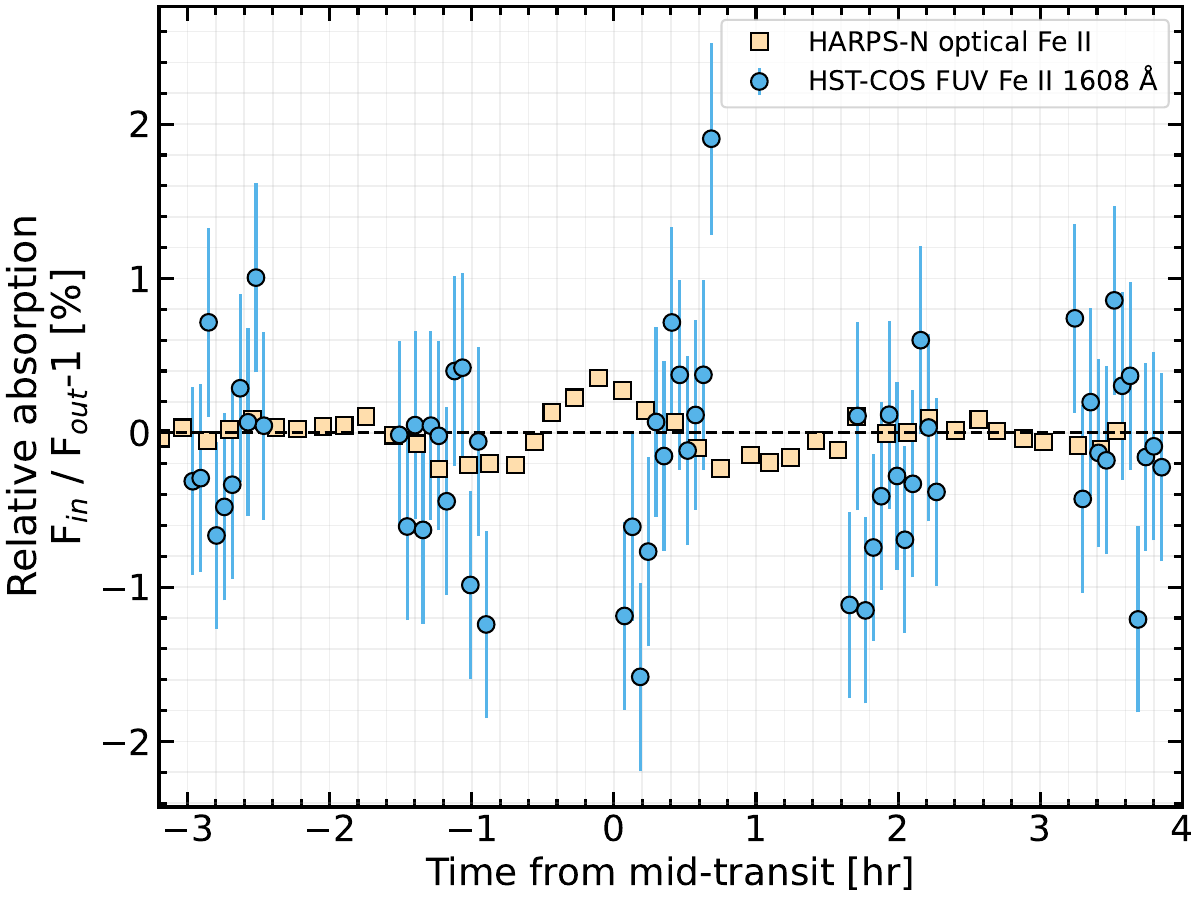}
    \caption{Comparison of the FUV \ion{Fe}{2} line (1608 \ang) and three coadded optical \ion{Fe}{2} lines (5018, 5169, 5316 \ang) from \citet{casasayas-barris_atmospheric_2019}. The light curves show the relative absorption of the line compared to the surrounding continuum. Beige squares show the HARPS-N optical light curve and blue circles show the COS FUV light curve. The optical light curve is shown before applying corrections for the RM and CLV effects, resulting in a prominent ``W'' shape which does not appear in the FUV light curve.}
    \label{fig:RM-FeII-compare}
\end{figure}

The Rossiter-McLaughlin (RM) effect and center-to-limb variation (CLV) are known to influence the shape of exoplanet transits. The RM effect occurs due to asymmetric coverage of the rotating stellar disk during the planet transit \citep{rossiter_detection_1924,mclaughlin_results_1924}. While the planet is not in transit, stellar lines are symmetrically broadened due to stellar rotation. Assuming an aligned orbit, the planet blocks some of the blueshifted light between ingress and transit center and redshifted light between transit center and egress. The result is a phase-dependent deformation of the radial velocity profile which can effect the shape of the observed transit light curve, typically manifesting as a smaller depth near mid-transit. CLV effects occur because of differences in stellar limb darkening within narrow wavelength regions. The intensity profile in the wings of an absorption feature may exhibit different behavior than within the core of the line; this has been seen in \ion{Ca}{2} H\&K and \ion{Na}{1} D lines in the Sun and as well as in exoplanet transits \citep{charbonneau_detection_2002,sing_hubble_2008,yan_effect_2017,czesla_center--limb_2015}. The effect of CLV on the transit light curve is similar to that of the RM effect, resulting in a ``W'' shaped transit with a smaller depth near mid-transit. The impacts of RM and CLV effects on the shape of the transit light curve are discussed in more detail in \S \ref{sec:chromosphere}. RM and CLV effects are significant in the optical observations of KELT-20 b and must be removed before performing the cross-correlation techniques commonly used in high resolution, ground-based spectroscopic analyses \citep{casasayas-barris_atmospheric_2019,hoeijmakers_high-resolution_2020,nugroho_searching_2020,bello-arufe_exoplanet_2022}. However, the strength of RM and CLV effects decrease with increasing stellar temperature and increasingly broad wavelength bands \citep{czesla_center--limb_2015}. Upon inspection---with the exception of \ion{Al}{2} and \ion{Si}{2}---we do not find evidence of the characteristic shapes from RM or CLV effects in either our broadband light curves (160 \ang\ width) or narrow absorption line light curves ($\sim$3 \ang\ width). For comparison, in Figure \ref{fig:RM-FeII-compare} we show the transit light curves of \ion{Fe}{2} in the optical \citep{casasayas-barris_atmospheric_2019} and in the FUV (this work). Considering the high temperature of KELT-20 and the relatively broad wavelength bands we use in our analysis, we assume that RM and CLV effects are negligible in our observations. Because the RM and CLV effects act to reduce the depth at mid-transit, there is a concern that including them in the transit model could artificially increase the transit depth by oversubtracting features which may not be present; because of this, we opt not to include these effects in our transit models. However, we note that excluding these effects from our model may impact the results of our fits, particularly for \ion{Al}{2} and \ion{Si}{2}, and that further observations are required to get complete phase coverage of the transit at high spectral resolution to be able to confidently assess the impact of RM and CLV effects in the FUV.

\subsection{Broadband analysis}
\label{subsec:broadband-transit}
We first consider the two broadband regions 1410--1570 \ang\ and 1600--1760 \ang. Wavelengths predicted to have strong atomic absorption features (i.e., \ion{Al}{2}, \ion{Si}{2}, \ion{C}{1}) are removed from the data before creating the light curves to avoid artificially increasing the broadband transit depth. The light curves are then binned to a 200 s cadence and normalized to the average rate in the final 1 ks of orbit 1 and all of orbit 5.

\subsection{Spectral analysis}
\label{subsect:wavelength-dependent}
To measure the spectral dependence of the planet radius over the full G160M wavelength band, we bin the events to a 300 s cadence and create 10 \ang\ wide spectral bins. The temporal binsize is increased to compensate for the lower signal within small wavelength bins. We use the same transit modeling procedure described in \S \ref{sec:model} and retrieve new LDCs for each bin using \texttt{limb-darkening}. For the spectral analysis, we do not exclude any atomic absorption features.

\subsection{Atomic absorption}
\label{sec:line-abs}
We searched for excess absorption in resonant transitions of each of the following species: \ion{C}{1} ($\lambda 1561$ \ang, $\lambda1657$ \ang), \ion{N}{1} ($\lambda\lambda1493,1495$ \ang), \ion{S}{1} ($\lambda1474$ \ang), \ion{Si}{2} ($\lambda\lambda1527,1533$ \ang), \ion{Al}{2} ($\lambda1671$ \ang), and \ion{Fe}{2} ($\lambda1608$ \ang). For the \ion{Fe}{2}, \ion{Al}{2}, and \ion{Si}{2} lines, we isolate a region $\pm300$ km s$^{-1}$ ($\sim3$ \ang\ full width) around the line center. Both \ion{C}{1} regions are broad multiplets containing several non-resolved lines which we sum over 1559.5--1562.5 \ang\ and 1655.5--1658.5 \ang, informed by the predicted width from the \citet{fossati_gaps_2023} model. The \ion{N}{1} doublet is closely spaced so we use a single $\pm300$ km s$^{-1}$ wide region centered between the lines to encompass them both. For \ion{S}{1} we use a smaller $\pm150$ km s$^{-1}$ ($\sim1.5$ \ang\ full width) wide region to avoid contribution from nearby non-S lines. The \ion{Fe}{2} line may contain contribution from nearby non-resonant \ion{Fe}{2} transitions. Each light curve is then binned to a 300 s cadence and fit using the same method as the broadband and spectral analyses.

\section{Results}
\label{sec:results}

\subsection{Transit depth}
\label{sec:results-depth}
\begin{figure}[!ht]
    \centering
    \includegraphics[width=0.7\linewidth]{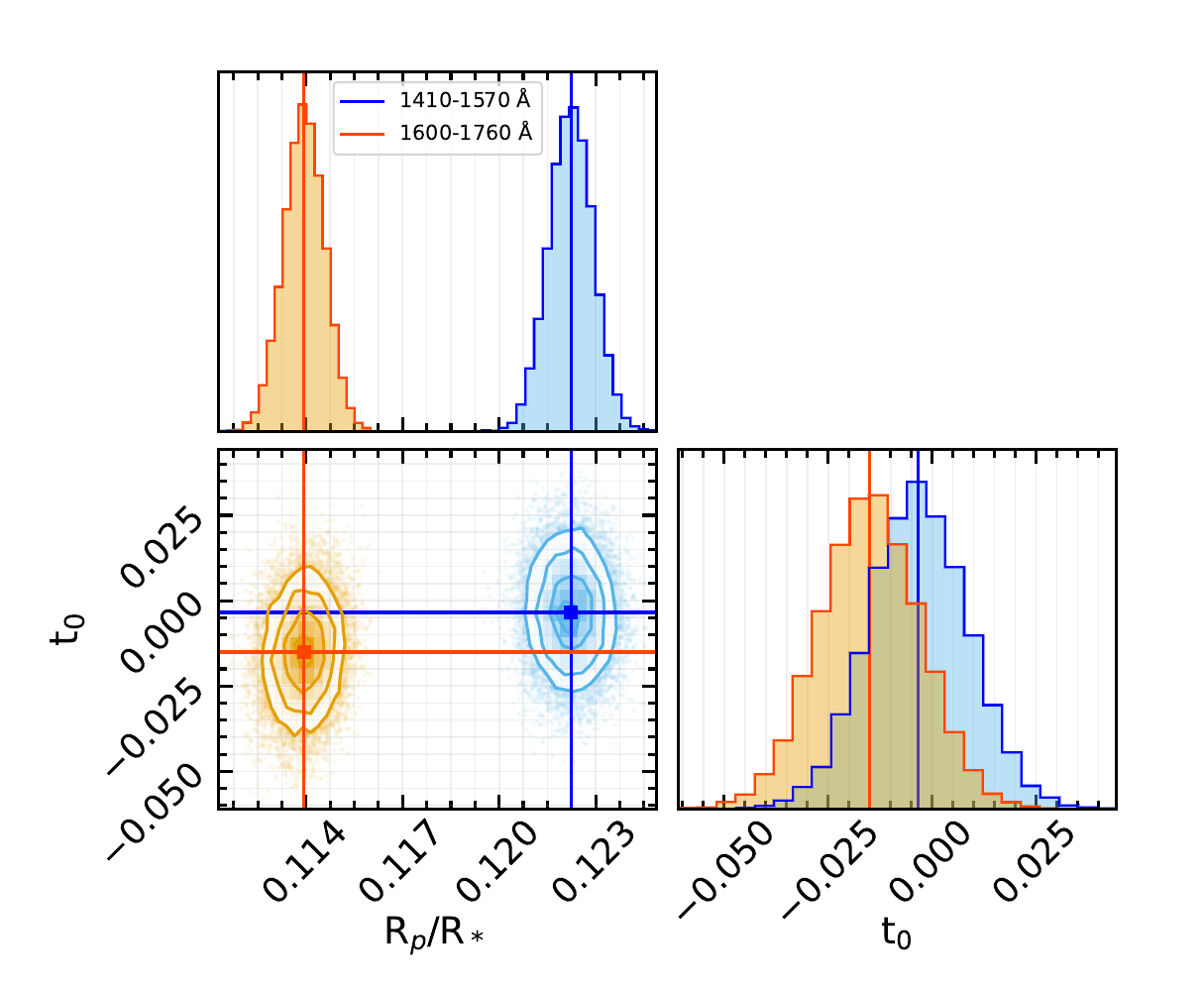}
    \caption{Corner plot showing posterior distributions for the broadband planet to star radius ratio ($R_p/R_*$) and time of mid-transit ($t_0$) retrieved from \texttt{PyTransit}. Blue distributions represent wavelengths 1410--1570 $\text{\AA}$ and orange distributions  wavelengths 1600--1760 $\text{\AA}$. Time of mid-transit ($t_0$) is shown in hours.}
    \label{fig:broadband-corner}
\end{figure}

\begin{figure}[!ht]
    \centering
    \includegraphics[width=\linewidth]{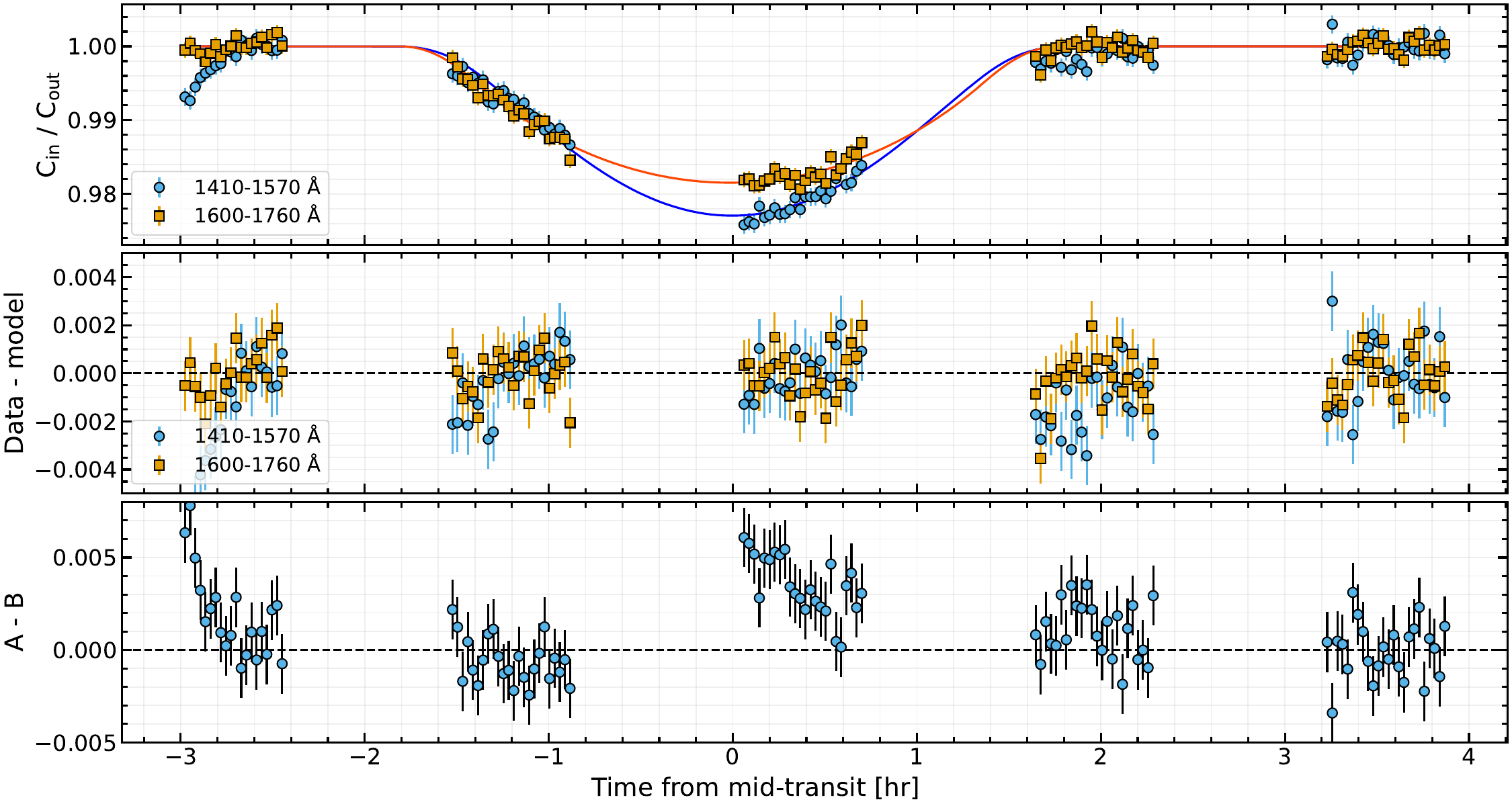}
    \caption{\textbf{Top:} Broadband transit light curve fit with the limb-darkened transit model described in \S \ref{sec:limb-darkening}. Blue circles represent wavelengths 1410--1570 \ang\ and orange squares 1600--1760 \ang. Solid lines show the respective best-fit model from \texttt{PyTransit}. \textbf{Middle:} Residual of data and model fits. Marker shapes and colors are the same as in the top panel. \textbf{Bottom:} Difference between segment A and B. Data are binned to 100 s.}
    \label{fig:broadband-transit}
\end{figure}

Figure \ref{fig:broadband-corner} shows the posterior distributions for $R_p/R_*$ and $t_0$ retrieved from our MCMC analysis and Figure \ref{fig:broadband-transit} shows the broadband transit light curve fit with the resulting transit model. The retrieved values are provided in Table \ref{tab:broadband-parameters}. The transit occurs at the expected time within the uncertainties, confirming previous results of no transit timing variations \citep{ivshina_tess_2022}, and we see no significant difference in the time of mid-transit between the two broadband regions. The \texttt{PyTransit} model shows good agreement with the observed data through ingress (orbit 2) and mid-transit (orbit 3) for both segments but over-predicts segment B during the end of egress (first half of orbit 4). We find a transit depth of $1.835\pm0.035$\% in segment A (1600--1760 \ang) and $2.296\pm0.042$\% in segment B (1410-1570 \ang), corresponding to $R_p/R_*$ of $0.11382\pm0.00064$ and $0.12202\pm0.00069$, respectively. The weighted average NUV transit depth from CUTE visits 2--4 is $2.337\pm0.242$\%.

\begin{table}[!ht]
\caption{Best fit values for broadband transit fit parameters retrieved from \texttt{PyTransit}}
\label{tab:broadband-parameters}
    \centering
    \begin{tabular}{CcC}
        \hline
        \hline
        \multicolumn{3}{c}{1600-1760 \ang}\\
        \hline
        \text{Parameter}   &   Description &   \text{Value}\\
        \hline
        R_p/R_* &   Planet-to-star radius ratio &   0.11383\pm0.00064\\
        t_0   & Mid-transit time relative to expected value [hr]  &   -0.019\pm0.012\\
        \hline\\
        \multicolumn{3}{c}{1400-1570 \ang}\\
        \hline
        \text{Parameter}   &   Description &   \text{Value}\\
        \hline
        R_p/R_* &   Planet-to-star radius ratio &   0.12202\pm0.00069\\
        t_0   & Mid-transit time relative to expected value [hr]  &   -0.017\pm0.012\\
        \end{tabular}
\end{table}

\subsection{Comparison to previous observations}
\label{sec:previous-obs}
\begin{figure}[!ht]
    \centering
    \includegraphics[width=1\linewidth]{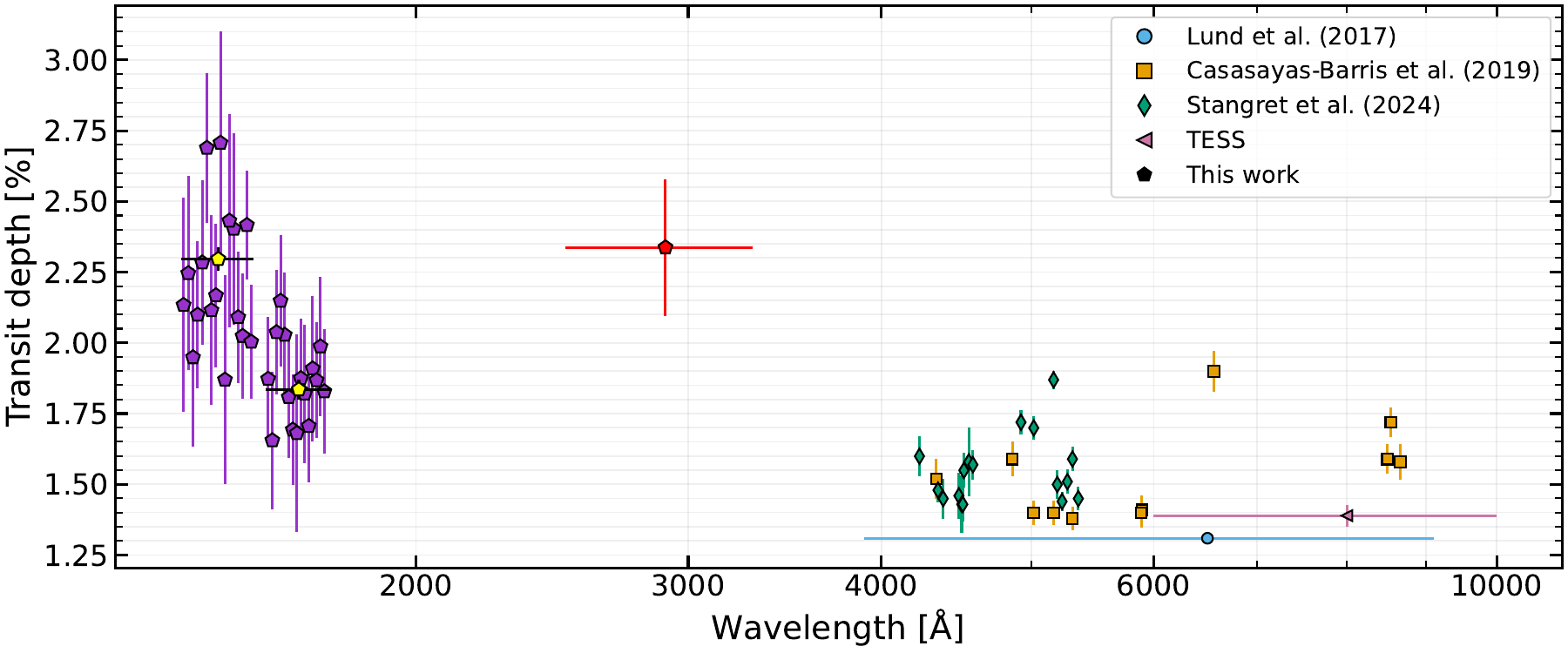}
    \caption{Transit depth spectrum of KELT-20 b in the FUV through near-IR. Results from this work are shown with pentagon markers and are further divided by color: purple markers represent the COS 10 \ang\ bins shown in Figure \ref{fig:wave-abs}, yellow represent the COS broadband segment A and B depths, and red represents the CUTE NUV transit depth. The \citet{lund_kelt-20b_2017} depth (circle marker) represents the photometric transit depth between 3900--9100 \ang. The \citet{stangret_gaps_2024} data (diamond markers) are measurements of  several \ion{Fe}{2} lines. The \citet{casasayas-barris_atmospheric_2019} data (square markers) are measurements of H-$\alpha$,$\beta$,$\gamma$,  \ion{Na}{1}, \ion{Mg}{1}, \ion{Fe}{2}, and \ion{Ca}{2}. The TESS transit depth (left-pointing triangle) is from the phase-folded light curve shown in Figure \ref{fig:light-curve}.}
    \label{fig:transit-spectrum}
\end{figure}

Here we present our FUV and NUV results in the context of previous observations from optical through near-IR. Figure \ref{fig:transit-spectrum} shows the transit depth of KELT-20 b from 1410--9100 \ang. The NUV depth from 2500--3300 \ang\ is the weighted average value from the CUTE visits 2--4. The photometric data show a large increase in transit depth between 3300--3900 \ang. This trend is expected, as UV photons are absorbed higher in the atmosphere, and has been seen in previous NUV observations of hot-Jupiters and UHJs \citep[e.g., WASP-121 b, WASP-178 b, HD 189733 b;][]{sing_hubble_2019,lothringer_uv_2022,cubillos_hubble_2023}.

Despite the transit depth being $\sim43$ \% deeper than the optical transit depth, when fit with \texttt{PyTransit} to retrieve the radius, the segment A (1600--1760 \ang) R$_p$/R$_*$ ratio is consistent with the published optical value \citep{lund_kelt-20b_2017,patel_empirical_2022,kokori_exoclock_2023}, potentially due to the steeper and more extreme limb darkening in the FUV compared to the optical.

\subsection{Comparison to atmospheric model}
\label{sec:model-comparison}
\begin{figure}[!ht]
    \centering
    \includegraphics[width=\linewidth]{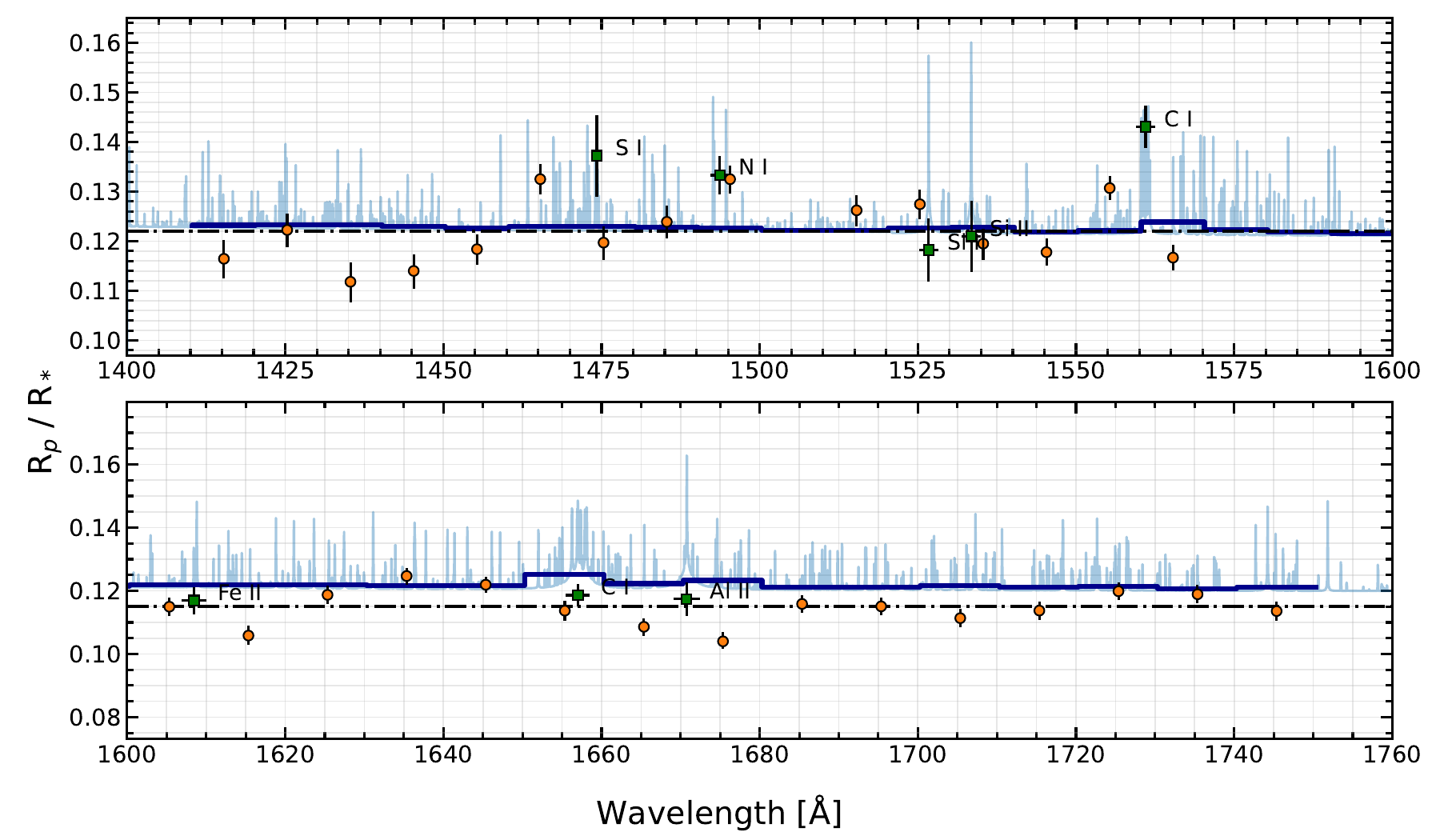}
    \caption{Comparison between the \citet{fossati_gaps_2023} NLTE transmission spectrum and the observed COS transit spectrum. The light blue spectrum is the synthetic model convolved with a small boxcar kernel for visibility. The solid dark blue line is the synthetic model binned to the same 10 \ang\ bins as the COS spectrum. Horizontal dash-dotted lines, circle markers, and square markers represent the radii retrieved from \texttt{PyTransit}. Dash-dotted lines represent the broadband values, circle markers the 10 \ang\ wide spectral bins, and square markers the atomic absorption lines discussed in \S \ref{sec:line-abs}.}
    \label{fig:wave-abs}
\end{figure}

\citet{fossati_gaps_2023} modeled the upper atmosphere of KELT-20 b using \texttt{CLOUDY} \citep{chatzikos_2023_cloudy} including non-LTE effects and were able to accurately reproduce the observed \ion{H}{1} Balmer line profiles. Figure \ref{fig:wave-abs} shows the NLTE atmosphere model extended to the FUV with our COS transit data binned to 10 \ang. At wavelengths shorter than $\sim1500$ \ang\ the continuum is dominated by the broad wings of H Lyman-$\alpha$ and Rayleigh scattering while longer wavelengths are dominated primarily by Fe absorption \citep{sing_hubble_2019,fossati_gaps_2023,petz_pepsi_2024}.

The NLTE model predicts a continuum level of $R_p/R_*\sim0.1228$\% in segment B and $R_p/R_*\sim0.1215$\% in segment A. Our segment B measurement is in agreement with the predicted model at the $1\sigma$ level. However, the 1600--1760 \ang\ region is unexpectedly low, falling $12\sigma$ below the theoretical model value. We currently do not have a compelling explanation for the discrepancy with the model, which has been able to replicate the optical observations as well as correctly predict the broadband radius in the segment B FUV region.

\subsubsection{Molecular absorption}
\label{sec:molecules}
\begin{figure}[!ht]
    \centering
    \includegraphics[width=\linewidth]{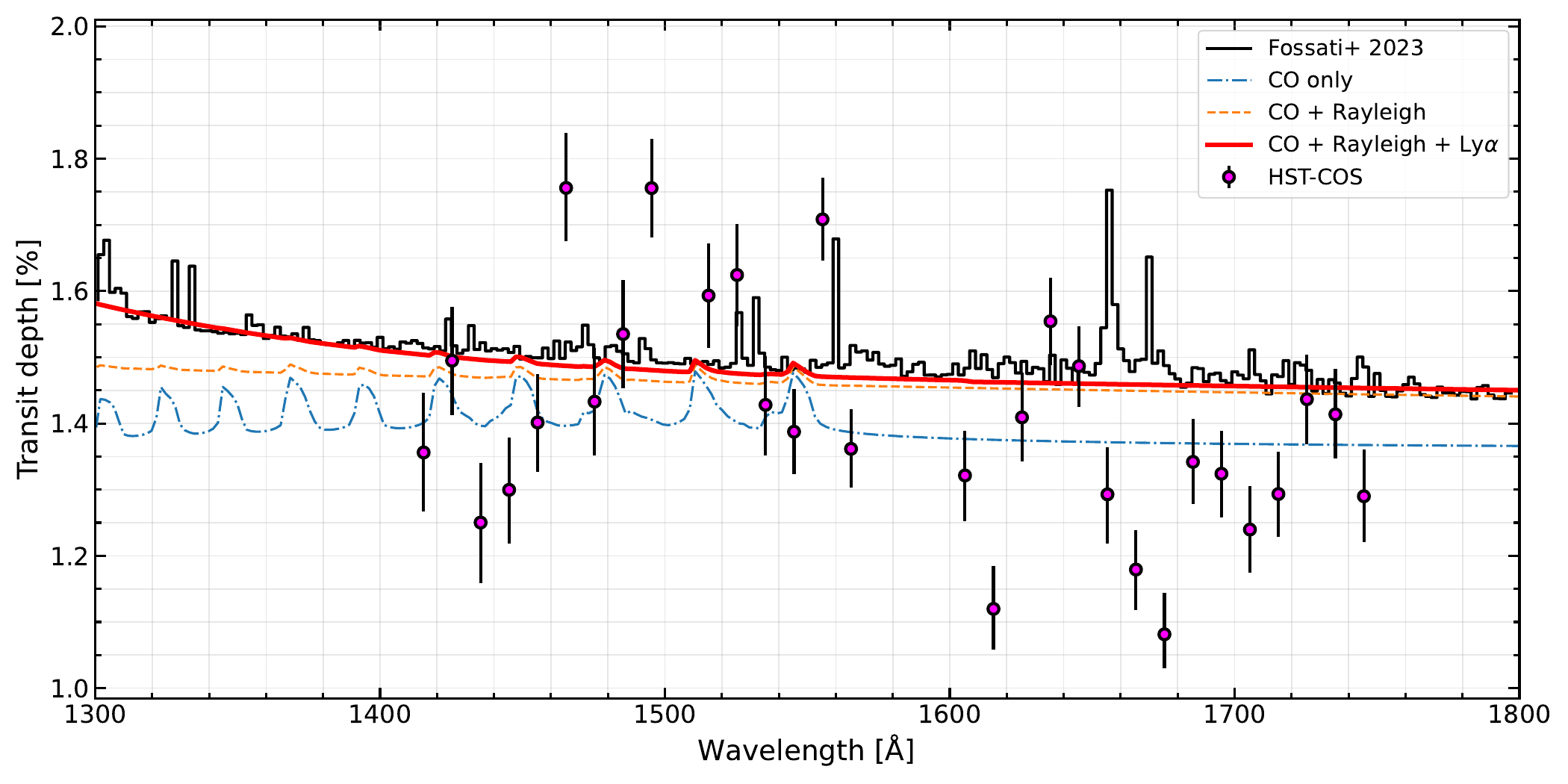}
    \caption{Synthetic transmission spectra including various sets of opacities. The dash-dotted blue line includes only CO. The dashed orange line includes CO, H, H$_2$, He, and a Rayleigh scattering component. The solid red line includes all of the previous plus \ion{H}{1} Ly$\alpha$. The solid black line is the \citet{fossati_gaps_2023} NLTE model binned to the same resolution as the CO models ($\sim2$ \ang\ per bin). The NLTE model is provided in units of R$_p$/R$_*$ and has been converted to transit depth according to the simplified relation $D=(R_p/R_*)^2$, which we note does not account for stellar limb-darkening. Purple circle markers represent our HST-COS observations.}
    \label{fig:CO-models}
\end{figure}

We first investigated the potential for molecular absorption to explain the difference in absorption depth between segments A and B. Most molecules in the upper atmosphere of UHJs are expected to be dissociated on the high temperature dayside. However, due to their high molecular bond strengths and ability to self-shield against photodissociation, H$_2$ and CO are able to survive temperatures of several thousand Kelvin and may remain intact. These molecules may thus show up in emission in the IR and absorption in the FUV \citep{lothringer_extremely_2018,fossati_extreme-ultraviolet_2018}, potentially explaining the difference in absorption depth seen between our two broadband regions (see Figure \ref{fig:CO-models}).

We first address the possibility of absorption from H$_2$. While H$_2$ absorption is typically strongest at wavelengths shorter than \ion{H}{1} Ly$\alpha$, at temperatures $2500<T<4000$ K it can also extend into the G160M bandpass. The \citet{fossati_gaps_2023} model includes H$_2$ and does not predict any strong H$_2$ absorption between 1400--1800 \ang. It is possible that the atmospheric heights probed in the FUV are higher than where H$_2$ remains abundant enough to be detectable. To address this, we modify the T-P profile by altering the amount of Fe and Mg in the atmosphere such that it remains cool enough to maintain high populations of H$_2$ at pressures down to 0.1 mbar while maintaining a high enough temperature in the upper atmosphere to still match the Balmer lines. Even with this highly modified T-P profile, H$_2$ absorption is not significantly predicted in the synthetic transmission spectrum.

Next we investigated potential CO absorption. Secondary eclipse observations in the IR have shown CO in emission in the atmosphere of KELT-20 b \citep{fu_strong_2022}. The CO A$^1\Pi$-X$^1\Sigma^+$ absorption bands occurs between $1270\lesssim\lambda\lesssim1720$ \ang\ and is strongest between $1400\lesssim\lambda\lesssim1560$ \ang\ \citep{watanabe_absorption_1953}. \citet{fossati_extreme-ultraviolet_2018} suggest that these FUV bands may be detectable during primary transit of stars with high photospheric FUV flux. CO is not included in the \citet{fossati_gaps_2023} model; to investigate the influence of CO, we ran a new radiative transfer model using \texttt{Pyrat Bay} \citep{cubillos_2021_pyrat}, adopting the same T-P profile as from \citet{fossati_gaps_2023} using three different sets of opacities: one including only CO \citep[derived from][]{france_co_2011}; one including CO plus H, He, and H$_2$ with Rayleigh scattering; and one with all of the previous opacities plus \ion{H}{1} Ly$\alpha$. The three models are shown relative to the \citet{fossati_gaps_2023} model in Figure \ref{fig:CO-models}. Note that the continuum shape is slightly different in the CO models because the radiative transfer code used assumed LTE while the \citet{fossati_gaps_2023} models include NLTE effects which alter the continuum level by overpopulating \ion{Fe}{2} and underpopulating \ion{Mg}{2} relative to the LTE models. Nonetheless, we consider the LTE CO models sufficient to draw first order conclusions regarding the effect of CO on the transmission spectrum.

While the CO features are very prominent in isolation, they are rapidly muted relative to the continuum when the Rayleigh and Ly$\alpha$ opacities are introduced and they are not observed in our spectrum. We therefore also rule out absorption by CO and conclude that the difference in broadband transit depth cannot be attributed to molecular H$_2$ or CO absorption.

\subsubsection{Atomic species}
\label{sec:results-atomic}
Light curves and transit models for each atomic species are shown in Figure 14. Transit depths and the corresponding R$_p$/R$_*$ and $t_0$ are reported in Table \ref{tab:spectral-radius}. The observed transit depth is the average value of the 3 deepest points during orbit 3. The reported values and uncertainties of R$_p$/R$_*$ and $t_0$ are the mean and $1\sigma$ standard deviation of the MCMC posterior distribution. We find tentative detections of excess absorption in \ion{Fe}{2} at $2.8\sigma$, \ion{N}{1} at $2.2\sigma$, and \ion{C}{1} ($\lambda1561$) at $2.0\sigma$.

\begin{deluxetable}{cCCCCCC}
\label{tab:spectral-radius}
\tablewidth{\textwidth}
\tablecaption{KELT-20 b transit depth and radius of atomic species}
\tablehead{\colhead{Name} & \colhead{Wavelength [\ang]} & \colhead{Observed Transit Depth [\%]} &   \colhead{Excess depth [\%]}    &   \colhead{R$_p$/R$_*$ [\%]}   &   \colhead{Excess R$_p$/R$_*$ [\%]}  &   \colhead{t$_0$ [hr]}}
\startdata
    Broad A &   1600-1760    &   1.835\pm0.035    &   ---   &   11.383\pm0.064  &   --- &   -0.019\pm0.012\\
    Broad B &   1410-1570   &   2.296\pm0.042 &   ---   &   12.202\pm0.069  &   --- &   -0.017\pm0.012\\
    \hline
    \ion{Al}{2} &   1670.79  &   2.386\pm0.368    &  0.551\pm0.370 & 11.738\pm0.532  &   0.355\pm0.536   &   0.044\pm0.097\\
    \ion{C}{1}\tablenotemark{a} &  1657.00    &   1.768\pm0.395  &  -0.066\pm0.397 &   11.861\pm0.365  &   0.479\pm0.37    &   -0.208\pm0.065\\
    \ion{Fe}{2} &   1608.45 &   2.692\pm0.300 &   0.858\pm0.302 &   11.689\pm0.434  &   0.307\pm0.438   &   -0.108\pm0.078\\
    \ion{C}{1}\tablenotemark{a}  &   1561.00    &   2.687\pm0.416   &   0.853\pm0.417   &   14.308\pm0.431  &   2.925\pm0.436   &   -0.064\pm0.064\\
    \ion{Si}{2} &   1533.45 &   2.696\pm0.514 &   0.401\pm0.516 &   12.101\pm0.721  &   -0.101\pm0.724  &   0.072\pm0.118\\
    \ion{Si}{2} &   1526.72 &   2.774\pm0.446 &   0.478\pm0.448 &   11.819\pm0.641  &   -0.383\pm0.0645 &   -0.209\pm0.102\\
    \ion{N}{1}\tablenotemark{a}  &   1493.70 &   2.966\pm0.303 &   0.670\pm0.306   &   13.331\pm0.389  &   1.13\pm0.395    &   -0.170\pm0.075\\
    \ion{S}{1}\tablenotemark{a}  &   1474.4 &   2.992\pm0.066   &   0.697\pm0.661    &   13.721\pm0.823  &   1.519\pm0.826   &     0.032\pm0.117\\
\enddata
\tablenotetext{a}{Includes multiple lines. Reported wavelength is the midpoint of the wavelength bin used in calculations.}
\end{deluxetable}

The high mass of KELT-20 b results in a large Roche lobe radius. Combined with the low expected XUV flux from the host star, it is predicted that the atmosphere is in a hydrostatic regime. Approximating the Roche lobe as in \citet{erkaev_roche_2007}:

\begin{equation}
    R_{rl} \approx a\left(\frac{q}{3}\right)^{1/3}
\end{equation}

where $q$ is the ratio of planet mass to star mass, $q=M_{pl}/M_{*}$, the Roche lobe is $R_{rl}/R_*\approx61.5$\%---many times larger than the radius of any of our detected atmospheric species. \citet{fossati_gaps_2023} calculated the Jeans parameter as a function of atmospheric pressure and found that it is greater than 20 in the middle and upper atmosphere. The large Jeans parameter and lack of observed metal species at high altitudes suggest that the atmosphere is not in a hydrodynamic escape regime.

\begin{figure}[!ht]
    \centering
    \includegraphics[width=\linewidth]{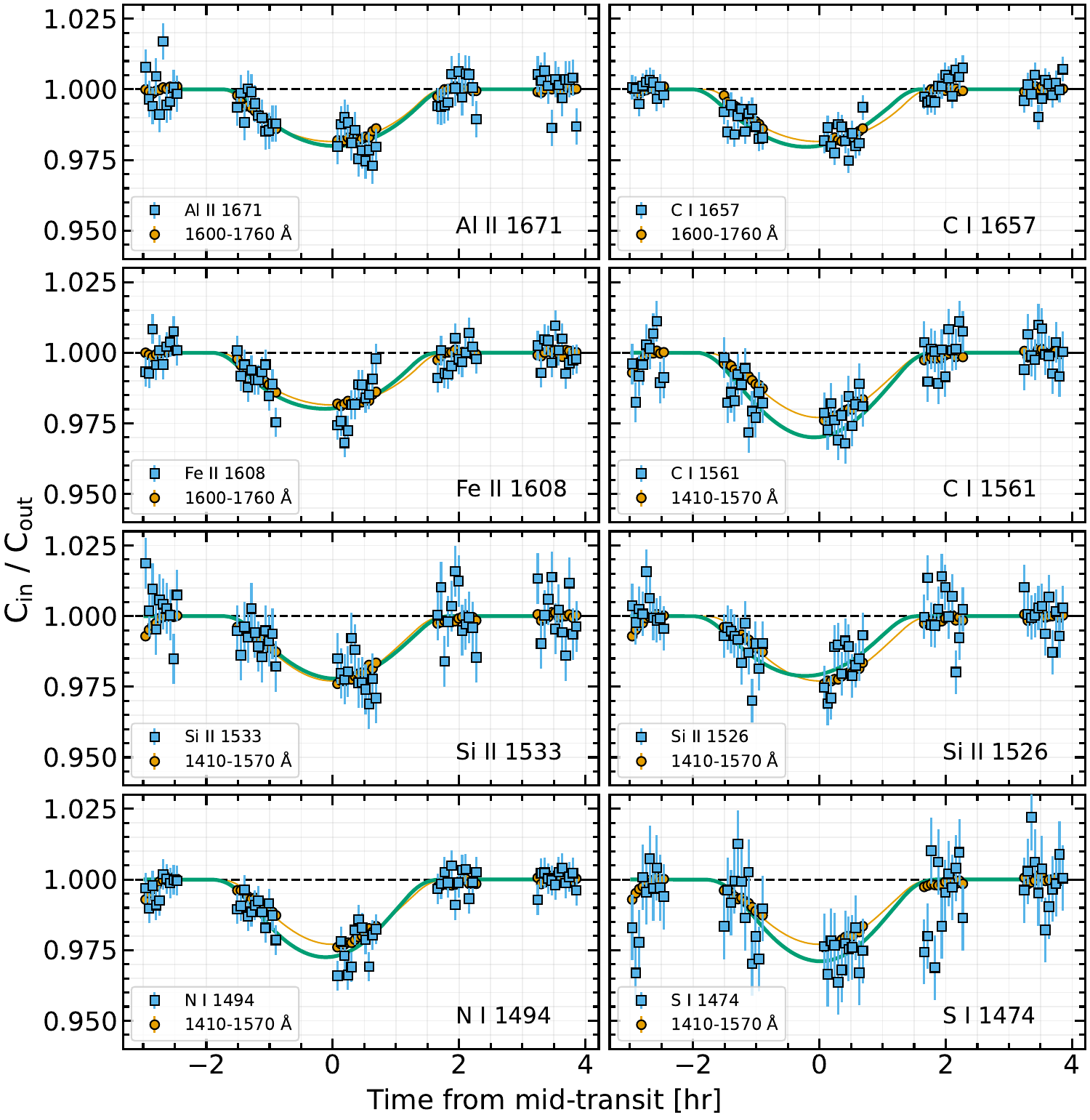}
    \caption{Light curves for each of the atomic species listed in Table \ref{tab:spectral-radius}. For each subplot, the blue squares represent the absorption line light curve and the orange circles represent the appropriate broadband region. The solid green line represents the \texttt{PyTransit} fit to the absorption line and the thin orange line represents the fit to the broadband region.}
    \label{fig:ion-lcs}
\end{figure}

\subsection{Stellar chromospheric activity}
\label{sec:chromosphere}
While we expect the stellar intensity profile to be limb darkened over broad wavelength bands, emission in narrow bands from chromospheric or coronal ions may result in a limb brightened profile. This behavior has been seen in \ion{Si}{4} in the Sun \citep{wiik_eruptive_1997,de_pontieu_new_2021,ayres_2021_solar}. In the context of a transiting planet, limb brightening would result in a ``W-shaped'' light curve, as the planet blocks proportionally more light at the bright edge of the disk than in the darker center \citep{assef_detecting_2009,schlawin_exoplanetary_2010,perdelwitz_influence_2025}.

\begin{figure}[!ht]
    \centering
    \includegraphics[width=\linewidth]{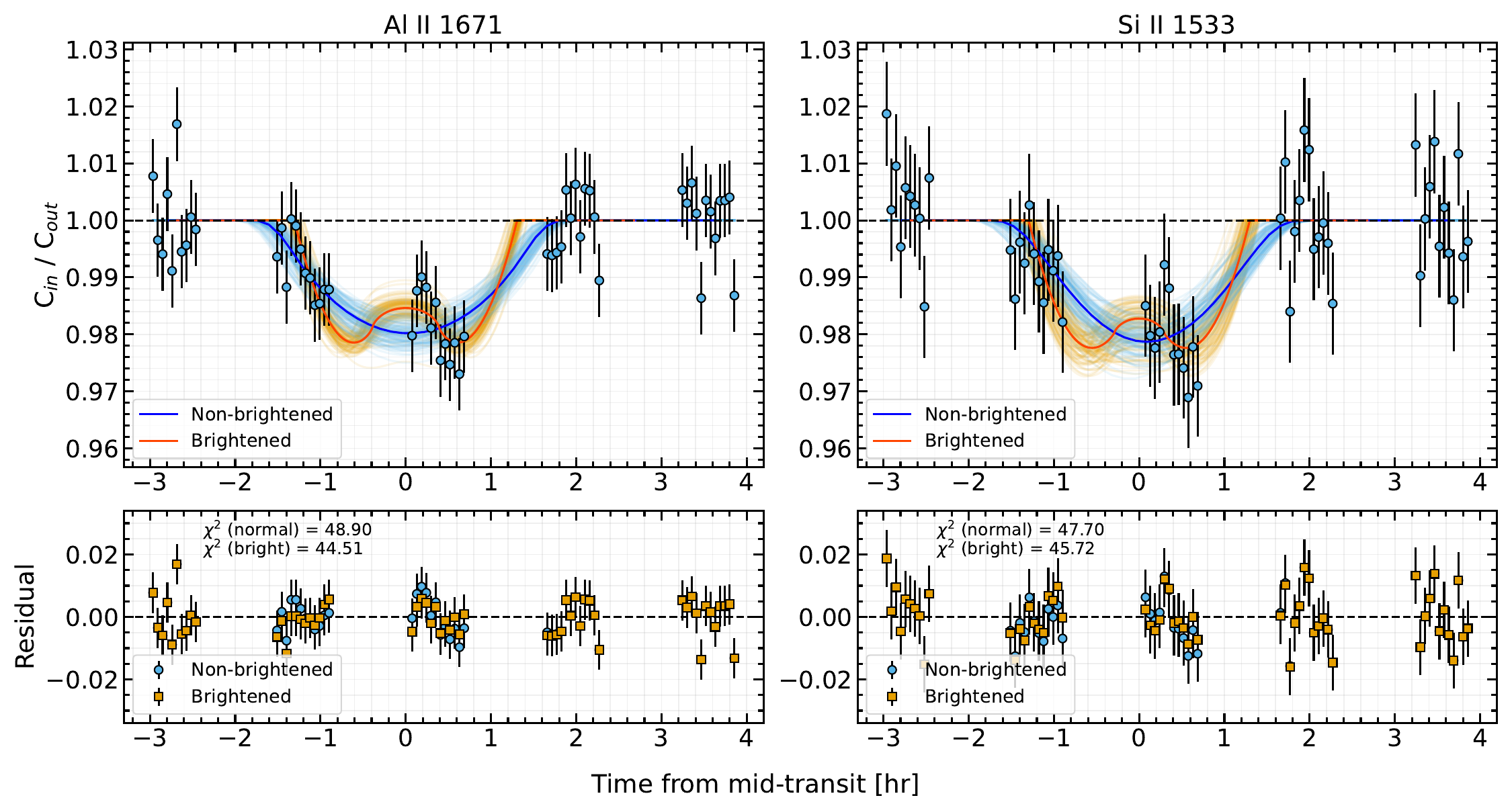}
    \caption{\textbf{Top:} Transit light curves of \ion{Al}{2} and \ion{Si}{2} binned to a 200 s cadence. Each transit has been fit with a ``normal'' limb darkened transit model as well as a ``W-shaped'' limb brightened model. Light colored model fits are randomly drawn samples from the MCMC trials. \textbf{Bottom:} Residuals between the COS observed transit and the two transit models.}
    \label{fig:limb-brightening}
\end{figure}

The transit light curves of \ion{Al}{2} ($\lambda1671$ \ang) and \ion{Si}{2} ($\lambda1533$ \ang) show tentative evidence of the W-shape characteristic of a limb brightened transit. Figure \ref{fig:limb-brightening} shows the transit light curves for \ion{Al}{2} and \ion{Si}{2} fit with \texttt{PyTransit} using a non-linear limb darkening law (see \S \ref{sec:limb-darkening}) as well as the limb brightened model described in \citet{schlawin_exoplanetary_2010}\footnote{Code for the \citet{schlawin_exoplanetary_2010} limb brightened transit model can be found at: \url{https://faculty.washington.edu/agol/chromosphere.html} (IDL) or \url{https://github.com/prbehr/chrom_exact} (python)}. The limb brightened model takes as parameters the ratio of planet-to-star radius and the transit impact parameter.

\begin{deluxetable}{ccC}[!h]
\label{tab:limb-brightened-fits}
\tablewidth{\textwidth}
\tablecaption{KELT-20 b limb brightened transit fits.}
\tablehead{\colhead{Parameter} & \colhead{Description} & \colhead{Retrieved value}}
\startdata
     & \ion{Al}{2}  &   \\
    \hline
    $R_p/R_*$   &   Planet-to-star radius ratio &   0.1222^{+0.0051}_{-0.0056}\\
    $b$   &   Impact parameter    &   0.8505^{+0.0116}_{-0.0139}\\
    $a$   &   Orbit semi-major axis [$R_*$]  &   12.5700^{+0.0116}_{-0.2059}\\
    $i$   &   Inclination [deg]   &   83.4178^{+0.1083}_{-0.0901}\\
    \hline
     &  \ion{Si}{2} &\\
    \hline
    $R_p/R_*$   &   Planet-to-star radius ratio &  0.1242^{+0.0076}_{-0.0209} \\
    $b$   &  Impact parameter &  0.8621^{+0.0267}_{-0.0209} \\
    $a$   &  Orbit semi-major axis [$R_*$] &  12.7409^{+0.3948}_{-0.3085} \\
    $i$   &  Inclination [deg] &  83.3277^{+0.1623}_{-0.2077} \\
\enddata
\tablecomments{Semi-major axis and inclination values represent the value required to obtain the best-fit impact parameter while keeping the other variable fixed to the published value.}
\end{deluxetable}

At face value, the limb brightened model seems to provide a better fit than the non-brightened fits; however, the retrieved impact parameter requires a dramatic increase from the published value of $b=0.503$ to $b=0.851$ for \ion{Al}{2} and $b=0.862$ for \ion{Si}{2}. At the extremes (i.e., changing \textit{only} the orbital separation or inclination), this would require an orbital separation of $a/R_*\sim12.6$ ($1.7\times$ larger than published value) or an inclination of $i\sim83$ deg (3 deg smaller than published value). Of course, it is likely that a change in impact parameter would come from some combination of smaller changes in both inclination and separation. The limb-brightened model also results in a much shorter transit duration than the non-brightened models, which agree with the duration of previous transit observations. Given the required deviations from the precise measurements of ground-based observations, we find a limb-brightened explanation unlikely.

On the other hand, CLV effects within absorption features can result in different limb darkening behavior in the wings of the line compared to the core. If the wings of the line show a less steep intensity profile, it can mimic the W-shaped limb-brightened profile because of the relative difference in intensity between the wings and the core \citep{czesla_center--limb_2015,reiners_solar_2023}. It is important to point out that the wings of the line are not actually limb-brightened and it is simply a relative effect. \citet{czesla_center--limb_2015} modeled CLV effects in the \ion{Na}{1} D$_1$ and D$_2$ lines for stellar effective temperatures of 4000, 5000, and 6000 K, over four wavelength bands with half-widths ranging from 0.1--1.5 \ang. The 6000 K, 1.5 \ang\ model, which is closest to our observations, showed a negligible contribution from CLV. Because the temperature of KELT-20 is $\sim2800$ K hotter, we expect the CLV contribution will be even smaller. However, \citet{yan_centre--limb_2015} also note that CLV effects will likely be more pronounced in the UV. Ultimately, without detailed modeling and full phase coverage observations, it is unclear how significant CLV effects are in our observations and we cannot rule it out as a contributing factor to the W-shaped light curves.

Additionally, \citet{casasayas-barris_atmospheric_2019} obtained high-resolution transit spectroscopy of KELT-20 b using HARPS-N (R=115,000) and CARMENES (R=94,600). After modeling and removing RM and CLV effects, their observations of H$\alpha$, \ion{Na}{1}, \ion{Mg}{1}, \ion{Fe}{2}, and \ion{Ca}{2} exhibit a residual W-shape similar to that seen in our \ion{Al}{2} and \ion{Si}{2} transits. They attribute the shape of the transit to incomplete removal of the RM effect resulting in a decreased depth near mid-transit.

Finally, we note that, to date, there have been no unambiguous detections of coronal or chromospheric emission from early A stars with $T_{\rm{eff}}\gtrsim8300$ K. Previous detections of X-ray emission and FUV chromospheric line emission (\ion{C}{3} and \ion{O}{6}) in early A stars have been attributed to unresolved low-mass companions which are X-ray luminous and support strong chromospheric activity \citep{simon_limits_2002,neff_o_2008,gunther_coronal_2022}. Considering the impact that RM and CLV effects, which we have not attempted to remove from our observations, may have on the transit shape, we conclude that the \ion{Al}{2} and \ion{Si}{2} transit shape likely stems from these effects. However, if the transit shape is indeed due to stellar limb-brightening, it would be indicative of chromospheric activity on a star thought to be too hot to support a chromosphere. Further FUV observations at shorter wavelengths to measure stellar chromospheric tracers such as \ion{C}{3}, \ion{O}{6}, and \ion{N}{5} could help provide a conclusive answer.

\section{Summary}
\label{sec:summary}

We have obtained the first FUV transit observations of an UHJ using HST-COS. KELT-20 b (MASCARA-2 b) orbits an A2 V star with $T_{\rm{eff}}\sim8800$ K which is FUV bright but thought to be weak in X-rays and EUV. Combined with its high mass, this makes KELT-20 b a unique target to investigate the atmospheric properties of a high mass planet in an FUV-dominated radiation environment. Results from our broadband and atomic FUV transit observations are summarized as follows:

\begin{enumerate}
    \item The broadband FUV transit depth of KELT-20 b increases with decreasing wavelength. The region between 1600--1760 \ang\ has a depth of $1.835\pm0.035$\% ($R_p/R_*=0.11383\pm0.00064\%$) and the region between 1400--1570 \ang\ a depth of $2.296\pm0.042$\% ($R_p/R_*=0.12202\pm0.00069\%$). The radius of the shorter wavelength segment is consistent with the model of \citet{fossati_gaps_2023} but the radius in the long wavelength segment is smaller than expected and is consistent with the optical radius. We are unable to provide an explanation for the difference in transit depth but rule out known instrumental systematics as well as atomic and molecular absorption (\S \ref{sec:results-depth}, \S \ref{sec:molecules}).
    \item We present tentative detections of excess absorption depth compared to the surrounding continuum in \ion{Fe}{2} ($\lambda1608$ \ang) at $2.8\sigma$,  \ion{N}{1} ($\lambda1474$ \ang) at $2.2\sigma$, and \ion{C}{1} ($\lambda1561$ \ang) at $2.0\sigma$. We do not find evidence of excess absorption in \ion{C}{1} ($\lambda1657$ \ang), \ion{Si}{2} ($\lambda\lambda1526,1533$ \ang), \ion{Al}{2} ($\lambda1671$ \ang), or \ion{S}{1} ($\lambda1474$ \ang) (\S \ref{sec:results-atomic}). All detected atomic species are well below the Roche lobe radius and we find no sign of hydrodynamic escape.
    \item \ion{Al}{2} and \ion{Si}{2} show ``W-shaped'' transits which could indicate stellar limb-brightening. However, the host star is not expected to have an active chromosphere, the limb-brightened transit fits require orbital parameters which are inconsistent with all previous observations, and our data reduction does not remove potential contamination from the Rossiter-McLaughlin effect or stellar center-to-limb variation. We conclude that the oddly shaped transits are likely the result of these unremoved astrophysical effects (\S \ref{sec:chromosphere}).
\end{enumerate}

\begin{acknowledgments}
\textit{Acknowledgments} We thank the anonymous referee of this report for their encouraging words and thoughtful suggestions. The HST observations presented here were acquired as part of the HST Cycle 30 program 17156, and supported by the associated grant to the University of Colorado, Boulder. The HST observations were obtained from the Mikulski Archive for Space Telescopes (MAST) at the Space Telescope Science Institute. The specific observations analyzed can be accessed via \dataset[doi:10.17909/ztc1-9v36]{https://doi.org/10.17909/ztc1-9v36}. The CUTE observations used in this work are archived at NExSci and can be accessed via \dataset[doi:10.26133/NEA43]{https://doi.org/10.26133/NEA43}.
\end{acknowledgments}

\vspace{5mm}
\facilities{HST-COS, CUTE}

\software{
    \texttt{Astropy} \citep{the_astropy_collaboration_astropy_2022},
    \texttt{SciPy} \citep{virtanen_scipy_2020},
    \texttt{NumPy} \citep{harris_array_2020},
    \texttt{Matplotlib} \citep{hunter_matplotlib_2007},
    \texttt{CLOUDY} \citep{chatzikos_2023_cloudy},
    \texttt{PyTransit} \citep{parviainen_pytransit_2015},
    \texttt{limb-darkening} ,\citep{espinoza_limb_2015},
    \texttt{Pyrat Bay} \citep{cubillos_2021_pyrat},
    \texttt{emcee} \citep{foreman-mackey_emcee_2013},
    \texttt{BATMAN} \citep{kreidberg_batman_2015},
    \texttt{CONTROL} \citep{sreejith_autonomous_2022}
}

%\appendix
\appendix
\section{Spectroscopic transit depth}

\begin{figure}[!h]
    \centering
    \includegraphics[width=0.8\linewidth]{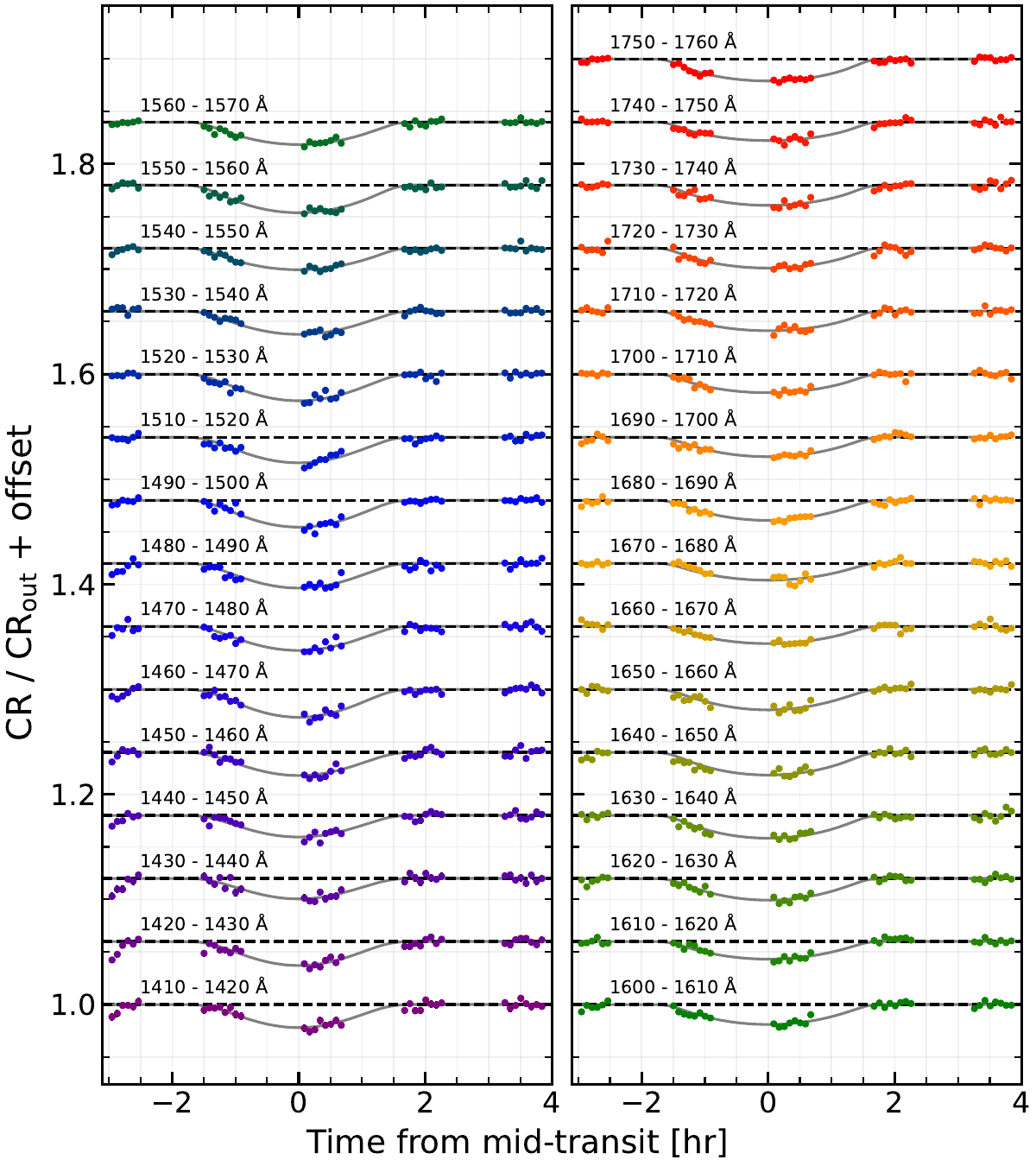}
    \caption{Spectroscopic light curves in 10 \ang\ bins at 300 s temporal cadence. \textbf{Left:} Light curves for bins contained in detector segment B. \textbf{Right:} Light curves for bins contained in detector segment A. Each panel is ordered with wavelength bins increasing along the y-axis and each light curve has been shifted by 0.06 for clarity. Wavelengths are color coded with purple being the shortest and red being the longest. Each light curve has been fit with a \texttt{PyTransit} model using the radius and LDCs provided in Table \ref{tab:spectroscopic-data}.}
    \label{fig:spec-light-curves}
\end{figure}

This Appendix provides a record of the spectroscopic data shown in Figure \ref{fig:wave-abs}.  Table \ref{tab:spectroscopic-data} tabulates the central wavelength, transit depth, and retrieved \texttt{PyTransit} values. Time of mid-transit has been fixed to zero for all fits. Emission lines are not tabulated here because they are recorded in Table \ref{tab:spectral-radius}. Figure \ref{fig:spec-light-curves} shows the transit light curve for each 10 \ang\ bin and the resulting \texttt{PyTransit} model fit.

\begin{deluxetable}{ccccccc}[!h]
\label{tab:spectroscopic-data}
\tablewidth{\textwidth}
\tablecaption{Record of spectroscopic data shown in Figure \ref{fig:wave-abs}. Transit depth is the average of the deepest three data points, radius is the value and $1\sigma$ error retrieved from \texttt{PyTransit}, and c$_i$ is the $i^{th}$ limb-darkening coefficient retrieved from \texttt{limb-darkening}.}
\tablehead{\colhead{Wavelength [\ang]} &   \colhead{Transit depth} &   \colhead{Radius}    &   \colhead{c$_1$} &   \colhead{c$_2$} &   \colhead{c$_3$} &   \colhead{c$_4$}}
\startdata
1415       & 0.0242$\pm$0.0020 & 0.1174$^{0.0034}_{0.0034}$ & -0.3361 & 2.3377  & -5.5987 & 4.5709  \\
1425       & 0.0242$\pm$0.0018 & 0.1197$^{0.0031}_{0.0031}$ & -0.3206 & 2.2690  & -5.4862 & 4.5129  \\
1435       & 0.0209$\pm$0.0020 & 0.1104$^{0.0037}_{0.0038}$ & -0.6304 & 3.5292  & -7.3856 & 5.4370  \\
1445       & 0.0241$\pm$0.0018 & 0.1140$^{0.0031}_{0.0032}$ & -0.1250 & 1.4373  & -4.1075 & 3.7838  \\
1455       & 0.0242$\pm$0.0016 & 0.1184$^{0.0028}_{0.0029}$ & -0.2007 & 1.7737  & -4.5814 & 3.9895  \\
1465       & 0.0284$\pm$0.0018 & 0.1303$^{0.0028}_{0.0028}$ & 0.0297  & 0.8379  & -3.1608 & 3.2927  \\
1475       & 0.0240$\pm$0.0018 & 0.1215$^{0.0030}_{0.0031}$ & 0.2905  & -0.2442 & -1.4493 & 2.4221  \\
1485       & 0.0231$\pm$0.0018 & 0.1231$^{0.0030}_{0.0030}$ & 0.0105  & 0.9292  & -3.2539 & 3.3109  \\
1495       & 0.0285$\pm$0.0016 & 0.1281$^{0.0026}_{0.0026}$ & 0.2706  & -0.1589 & -1.6081 & 2.5149  \\
1515       & 0.0269$\pm$0.0017 & 0.1252$^{0.0028}_{0.0028}$ & 0.1575  & 0.3620  & -2.4111 & 2.9004  \\
1525       & 0.0263$\pm$0.0017 & 0.1286$^{0.0027}_{0.0028}$ & 0.1231  & 0.4081  & -2.2140 & 2.6860  \\
1535       & 0.0231$\pm$0.0016 & 0.1211$^{0.0027}_{0.0028}$ & 0.2431  & -0.1468 & -1.1549 & 2.0650  \\
1545       & 0.0215$\pm$0.0014 & 0.1181$^{0.0024}_{0.0025}$ & 0.5117  & -1.7425 & 2.0447  & 0.2049  \\
1555       & 0.0264$\pm$0.0014 & 0.1336$^{0.0021}_{0.0021}$ & 0.5092  & -1.6474 & 1.8016  & 0.3567  \\
1565       & 0.0216$\pm$0.0013 & 0.1202$^{0.0022}_{0.0022}$ & 0.5937  & -1.8459 & 1.9325  & 0.3515  \\
1605       & 0.0204$\pm$0.0014 & 0.1146$^{0.0026}_{0.0026}$ & 0.5947  & -2.2205 & 3.2307  & -0.5825 \\
1615       & 0.0189$\pm$0.0014 & 0.1076$^{0.0027}_{0.0028}$ & 0.5065  & -1.7957 & 2.4837  & -0.1775 \\
1625       & 0.0228$\pm$0.0014 & 0.1190$^{0.0025}_{0.0025}$ & 0.6130  & -2.1291 & 2.8078  & -0.2631 \\
1635       & 0.0226$\pm$0.0014 & 0.1225$^{0.0023}_{0.0024}$ & 0.6432  & -2.3692 & 3.4478  & -0.6938 \\
1645       & 0.0224$\pm$0.0013 & 0.1226$^{0.0022}_{0.0022}$ & 0.6009  & -2.3302 & 3.6125  & -0.8607 \\
1655       & 0.0210$\pm$0.0016 & 0.1158$^{0.0028}_{0.0028}$ & 0.5072  & -1.4914 & 1.6034  & 0.4024  \\
1665       & 0.0169$\pm$0.0013 & 0.1069$^{0.0025}_{0.0025}$ & 0.6237  & -2.2915 & 3.3868  & -0.6937 \\
1675       & 0.0196$\pm$0.0015 & 0.1058$^{0.0028}_{0.0028}$ & 0.5880  & -2.1115 & 3.0660  & -0.5191 \\
1685       & 0.0202$\pm$0.0014 & 0.1153$^{0.0024}_{0.0025}$ & 0.5337  & -1.6313 & 2.0086  & 0.1083  \\
1695       & 0.0187$\pm$0.0014 & 0.1141$^{0.0025}_{0.0026}$ & 0.5892  & -2.0084 & 2.8606  & -0.4231 \\
1705       & 0.0183$\pm$0.0014 & 0.1123$^{0.0026}_{0.0027}$ & 0.4527  & -1.6621 & 2.8999  & -0.6864 \\
1715       & 0.0206$\pm$0.0014 & 0.1158$^{0.0025}_{0.0025}$ & 0.3971  & -1.5007 & 2.8310  & -0.7292 \\
1725       & 0.0199$\pm$0.0015 & 0.1168$^{0.0026}_{0.0026}$ & 0.4212  & -1.5528 & 2.8212  & -0.6895 \\
1735       & 0.0214$\pm$0.0015 & 0.1179$^{0.0026}_{0.0026}$ & 0.3928  & -1.3599 & 2.4329  & -0.4688 \\
1745       & 0.0199$\pm$0.0015 & 0.1132$^{0.0028}_{0.0028}$ & 0.3722  & -1.3720 & 2.7096  & -0.7159 \\
1755       & 0.0208$\pm$0.0014 & 0.1231$^{0.0024}_{0.0025}$ & 0.3970  & -1.4509 & 2.8044  & -0.7539
\enddata
\end{deluxetable}

\bibliography{KELT-20}{}
\bibliographystyle{aasjournal}

\end{document}